\documentstyle[12pt,epsf]{article}

\def \a   {\alpha}
\def \ab  {{\alpha \beta}}
\def \ah  {{1\over 2}}
\def \b   {\beta}
\def \be  {\begin{equation}}
\def \bea {\begin{eqnarray}}
\def \bl  {\bigl}
\def \bgl {\biggl}
\def \br  {\bigr}
\def \bgr {\biggr}

\def \cd  {{\gamma \delta}}
\def \cR  {{\cal R}}
\def \ct  {\cdot}

\def \d   {\partial}

\def \dt  {\delta}
\def \Dt  {\Delta}
\def \ee  {\end{equation}}
\def \eea {\end{eqnarray}}
\def \ep  {\epsilon}	
\def \2e  {2+\epsilon}
\def \et  {\eta}
\def \g   {\gamma}
\def \G   {\Gamma}
\def \h   {\hat}
\def \idx {\int d^D x}
\def \inf {\infty}

\def \k   {\kappa}

\def \lbl {\label}
\def \lts {\ldots}

\def \m   {\mu}
\def \mn  {{\mu \nu}}
\def \n   {\nu}
\def \nab {\nabla}
\def \nn  {\nonumber}

\def \ol  {\overline}
\def \ola {\overleftarrow}
\def \ov  {\over}
\def \p   {\phi}

\def \ps  {\psi}

\def \qq  {\qquad}
\def \r   {\rho}
\def \rgh {\sqrt{\hat g}}
\def \ra  {\rightarrow}
\def \rs  {{\rho \sigma}}
\def \s   {\sigma}
\def \sr  {\sqrt}

\def \t   {\tau}
\def \til {\tilde}

\def \vp  {\varphi}

\begin{document}

\setlength{\baselineskip}{7mm}

\begin{titlepage}
  \renewcommand{\thefootnote}{\fnsymbol{footnote}}
  \begin{normalsize}
    \begin{flushright}
      TIT-HEP-343 \\
      August 1996
    \end{flushright}
  \end{normalsize}
  \begin{Large}
    \vspace{2cm}
    \begin{center}
      {\LARGE Two--loop Prediction for Scaling Exponents in $(2 + \ep )$-dimensional 
              Quantum Gravity } \\
    \end{center}
  \end{Large}
\vspace{20mm}
\begin{center}
  Toshiaki A{\sc ida}\footnote 
  {E-mail address : aida@th.phys.titech.ac.jp} \\ 
  {\sc and}\\
  Yoshihisa K{\sc itazawa}\footnote
  {E-mail address : kitazawa@phys.titech.ac.jp} \\
\vspace{1cm}
  {\it Department of Physics, Tokyo Institute of Technology,} \\
  {\it Oh-okayama, Meguro-ku, Tokyo 152, Japan} \\
\vspace{1cm}
\end{center}

\hspace{1cm}
\begin{center}
\begin{abstract}
\setlength{\baselineskip}{7mm}
We perform the two loop level renormalization of quantum gravity
in $2+\ep$ dimensions. 
We work in the
background gauge whose manifest covariance
enables us to
use the short distance expansion of the Green's functions.
We explicitly show that the theory is renormalizable
to the two loop level in our formalism.
We further make a physical prediction for the scaling relation
between the gravitational coupling constant and the cosmological constant
which is expected to hold at the short distance fixed point
of the renormalization group.
It is found that the two loop level calculation is necessary
to determine the scaling exponent to the leading order in
$\ep$.
\hspace*{0.7\parindent}%
\end{abstract}
\end{center}
\end{titlepage}
\vfil\eject

\newpage

\section{Introduction}
\setcounter{equation}{0}
\lbl{sec:int}

We have developed the $2+\ep$ dimensional expansion of quantum gravity
extensively based on the pioneering works\cite{W,GKT,CD,KN}
and the exact solution of the two dimensional quantum gravity\cite{KPZ,D,DK}. 
It is well known that the conformal
invariance of the theory cannot be maintained in general due to
the conformal anomaly. However the structure of the conformal
anomaly is well understood in two dimensional quantum gravity 
and the conformal anomaly can be
cancelled by introducing the ``linear dilaton" type coupling\cite{CFMP,FT}.

We have proposed to introduce a ``linear dilaton" type coupling
in our formulation
which is chosen to cancel the conformal anomaly
in $2+\ep$ dimensions\cite{KKN1,KKN2,AKKN}.
We have formulated a systematic procedure to renormalize
the theory to all orders based on this idea.
We have also developed a formal proof of the
renormalizability within this scheme to all orders\cite{K,KKN3}. 
It is important to carry out an explicit calculation at the
two loop level to demonstrate the validity of this scheme.
Our previous work in this direction is \cite{AKNT} in
which we have computed the two loop counter terms to the leading order
in the central charge $c$.
In this paper, we complete the full two loop level renormalization of the
theory by determining the counter terms which render 
the effective action finite. 
We also renormalize the cosmological constant operator to the two
loop level.

We have adopted the background gauge formalism in our investigations.
We decompose the fields into the background fields and the quantum
fields. The background fields appear as the external fields and
the quantum fields appear in the loops. 
The background gauge has the advantage to maintain the manifest
covariance with respect to the background metric.
It enables us to use a manifestly covariant calculation procedure.
We determine the two loop level counter terms which makes 
the effective action finite. The effective action is
a functional of the background fields and hence manifestly
covariant. In order to achieve this goal, we also need to make the
two point functions of the quantum fields finite.
It is because they appear as the external fields for
the one loop subdiagrams of the two loop diagrams.
In this paper we report the result of these calculations.

In addition to demonstrate the validity of our renormalization scheme, 
there is a merit of these laborious calculations. 
They enable us to make a physical prediction
for the scaling relation which is supposed to
hold at the short distance fixed point of the renormalization
group. In our previous works, it is found that the anomalous
dimension of the cosmological constant operator is large
which almost make it to be a marginal operator at the short
distance fixed point of the renormalization group\cite{KKN1,AKKN}.
As it turns out that the two loop level calculation
enables us to calculate the scaling dimension of the
cosmological constant operator and it is found to be 
relevant.
It is because the scaling dimension is $O(\ep )$ and the two loop
level calculation is required to determine it.
We are then able to make a physical prediction for the scaling relation
between the gravitational coupling constant and the cosmological constant
at the ultraviolet stable fixed point of the renormalization group.
This prediction can be compared with the scaling exponent
which is measured in a
recent numerical simulation\cite{BKK}
of four dimensional quantum gravity by putting $\ep = 2$.

The organization of this paper is follows.
In section 2, we explain our calculation procedure which is
manifestly covariant with respect to the background metric.
We then apply the covariant calculation procedure
to the $2+\ep$ dimensional quantum gravity.
In section 3, we determine the counter terms 
up to the two loop level in the pure
Einstein gravity theory with no cosmological constant. 
In section 4, we renormalize the cosmological 
constant operator and calculate the scaling dimensions
of the relevant operators.
We make a physical prediction for the scaling relation
at the short distance fixed point of the renormalization group.
We conclude this paper in section 5.
The necessary information on the short distance singularities
of the products of the Green's functions is listed in the appendix A.

Our conventions for the geometric tensors are those of 
t'Hooft and Veltman\cite{tHV}.
The background metric is denoted by $\hat{g}_{\mu\nu}$ and 
the tensor indices are raised and lowered by the background metric.
The covariant derivatives should be understood to be taken with 
respect to the background metric.

%\newpage

\section{Covariant Calculation}
\setcounter{equation}{0}
\lbl{sec:cov}

In this section, we describe a covariant calculation method 
with respect to arbitrary backgrounds. We subsequently apply
it to the two-loop 
renormalization in $(2 + \ep )$-dimensional quantum gravity. 
The covariant method enables us to elucidate the general structure 
of the exact Green's functions in arbitrary Riemannian manifolds. 
Especially, we can express the short distance behavior of the Green's 
functions in a manifestly covariant way. This feature provides us 
a great help to perform multi-loop calculations.

%\vspace{1cm}

\subsection{Heat Kernel Methods}

Here, we illustrate the method by which we can 
express the short distance behavior the of exact propagators 
in a manifestly covariant way.
Although it is well explained in the literature
\cite{JO1,JO2}, we recapitulate it here for selfcontainedness. 

%\vspace{1cm}

We consider the elliptic differential operators defined 
on a $D$-dimensional Riemannian manifold of the form,
\be
\Dt_x \ = \ I \ \nab_x^2 \ + \ P(x) ,\ \ 
\nab^2 \ = \ \nab^\m \nab_\m,
\ee
Here $\nab_\m = \d_\m + \G_\m$ is the covariant derivative 
and $P$ denotes a generic matrix.
The exact propagator $G_\Dt$ is defined as the Green's function 
for the operator $\Dt$ in a manifestly covariant form.
\be
\Dt_x \ G_\Dt (x,x') \ = \ - I \ \dt^D (x,x'),
\lbl{eqn:defG}
\ee
We note that $G_\Dt (x,x')$ transforms as a bi-scalar or a 
bi-vector,$\lts$ according to the types of the fields on which 
the operator $\Dt$ acts. $\dt^D (x,x')$ is the bi-scalar 
$\dt$-function on a $D$-dimensional curved space satisfying
\be
\idx ' \sr{\h g '} \ \dt^D (x,x') \ f(x') \ = \ f(x).
\ee

To describe the short distance behavior of the exact propagator 
in a manifestly covariant way, we adopt the heat kernel method 
by De Witt \cite{DW}. The heat kernel for the operator $\Dt$ is 
defined with the initial condition by
\bea
& & \bgl( {\d \ov \d \t} - \Dt_x \bgr) \ 
{\cal G}_\Dt (x,x';\t) \ = \ 0 , \nn \\
& & \lim_{\t \ra +0} \ {\cal G}_\Dt (x,x';\t) \ = \ 
I \ \dt^D (x,x') .
\lbl{eqn:defhk}
\eea
We can easily write down the formal solution of the heat kernel.
\be
{\cal G}_\Dt (x,x';\t) \ = \ 
< x'| \ e^{\t \Dt } \ |x > \ . 
\lbl{eqn:fs1}
\ee

It is known that the heat kernel has the following asymptotic 
expansion in the small $\t$ limit.
\be
{\cal G}_\Dt (x,y;\t) = 
{v^\ah (x,x') \ov (4 \pi \t)^{D \ov 2}} \ 
e^{- {\s (x,x') \ov 2 \t}} \ 
\sum_{n=0}^\inf a^\Dt_n (x,x') \ \t^n .
\lbl{eqn:asyexp}
\ee
Here, $\s (x,x')$ is the bi-scalar geodetic interval defined to be equal to 
the half of the square of the geodetic distance between $x$ and $x'$. 
We shall call it the {\it geodetic interval}. We refer to a multi-component 
function of the two spacetime points, which transforms like the direct 
product of two scalars, as a bi-scalar function. In a similar way, we 
define bi-vector functions and bi-tensor ones. 
$\s (x,x')$ is a single-valued function for sufficiently close 
points $x,x'$. This is satisfied in our case since we only have 
to consider the short distance behavior to obtain ultraviolet divergences.

The bi-scalar of the geodetic interval is found to satisfy an important 
differential equation at an endpoint $x$. 
\be
\s (x,x') \ = \ \ah \ \h g^\mn (x) \s_\m \s_\n \ \ , \ \ 
\s_\m \ = \ {\d \ov \d x^\m} \s (x,x') .
\lbl{eqn:defs}
\ee
$v (x,x')$ is the bi-scalar Van Vleck-Morette 
determinant, defined by
\bea
v (x,x') & = & \h g^{- \ah} (x) \ \h g^{- \ah} (x') \ det(- \s_{\m \n '} ) ,
\nn \\
\s_{\m \n '} & = & {\d^2 \ov \d x^\m \d x'^{\n '} } \s (x,x') .
\eea
Hereafter, we often indicate the covariant differentiations 
of $\s$ and $v$ at $(x') \ x$ by using the (primed) subscripts.
$v$ satisfies the following equation which can be derived from the 
equation (\ref{eqn:defs}).
\be
v^\ah \nab^2 \s \ + \ 2 \s^\m \nab_\m v^\ah \ = \ 
D \ v^\ah .
\lbl{eqn:defd}
\ee
We note that $\s$ and $v$ depend only on the background metric 
locally, while the Seeley coefficients $a^\Dt_n$ are determined by 
the structure of the operator $\Dt$. 

The Seeley coefficients $a^\Dt_n (x,x')$ 
are defined inductively by the following equations.
\bea
\s^\m \ \nab_\m a^\Dt_0 & = & 0 \ , \ a^\Dt_0 | \ = \ I \ , 
\lbl{eqn:defSc1} \\
n \ a^\Dt_n \ + \ \s^\m \ \nab_\m a^\Dt_n & = & 
v^{-\ah} \ \Dt_x (v^\ah a^\Dt_{n-1}) ,
\lbl{eqn:defSc2}
\eea
which are obtained by substituting the asymptotic 
expansion (\ref{eqn:asyexp}) into the equations (\ref{eqn:defhk}).
Here we use a vertical bar to denote the coincidence limit 
of the two arguments: $x' = x$, {\it i.e.} $a_0 | = a_0 (x,x)$. 
%The definition (\ref{eqn:defSc1}) of the $0$-th Seeley 
%coefficient $a^\Dt_0$ requires the covariant 
%derivative of it along the geodesic to vanish. 
The equation (\ref{eqn:defSc1}) shows that
$a^\Dt_0 (x,x')$ is the two point matrix 
which makes the parallel displacement of the fields at $x'$ 
to the ones at $x$ along the geodesic connecting them. 
We also note that the $a_n$'s depend on both the metric and the operator 
$\Dt$ in a local fashion. 

The equations (\ref{eqn:defs}), (\ref{eqn:defd}), (\ref{eqn:defSc1}) 
and (\ref{eqn:defSc2}) 
play crucial roles when we obtain the coincidence limits 
of (the derivatives of) $\s$, $v$  and $a_n$. 
The coincidence limits of them are essential for our calculations 
which are based on the manifestly covariant expansions of the
Green's functions. 
Here, we tabulate 
the results of the coincidence limits of (the derivatives of)
$\s$ and $v$ to the order 
relevant to our calculation. 
\bea
\s | & = & 0 \ , \nn \\
\nab_\m \s | & = & 0 \ , \nn \\
\nab_\m \nab_\n \s | & = & \h g_\mn \ , \nn \\
\nab_\m \nab_\n \nab_\r \s | & = & 0 \ , \nn \\
\nab_\m \nab_\n \nab_\r \nab_\s \s | & = & 
- {1 \ov 3} (\h R_{\s \m \r \n } + \h R_{\s \n \r \m }) \ . 
\lbl{eqn:clim1a} \\
v^\ah | & = & 1 \ , \nn \\
\nab_\m v^\ah | & = & 0 \ , \nn \\
\nab_\m \nab_\n v^\ah | & = & - {1 \ov 6} \h R_\mn \ .
\lbl{eqn:clim1b}
\eea

Now, we can find the short distance behavior of the exact 
Green's function (\ref{eqn:defG}) in a manifestly covariant form 
through the relation which is obtained from 
(\ref{eqn:defG}) and (\ref{eqn:fs1}).
\bea
G_\Dt (x,x') 
& = & < x' | - {I \ov \Dt } | x > \ , \nn \\
& = & \int_0^\inf d \t \ {\cal G}_\Dt (x,x';\t) \ .
\eea
Corresponding to the short proper time expansion of the heat kernel 
(\ref{eqn:asyexp}), we obtain the short distance behavior of the 
Green's function as
\be
G_\Dt (x,x') \ = \ G_0 (x,x') \ a^\Dt_0 (x,x') \ + \ 
G_1 (x,x') \ a^\Dt_1 (x,x') \ + \ \ol G_\Dt (x,x') ,
\lbl{eqn:expG}
\ee
where
\bea
G_0 (x,x') & = & v^\ah (x,x') \ \bgl[ \
{\G ({\ep \ov 2}) \ov 4 \pi^{D \ov 2}} 
\ \bl( 2 \s (x,x') \br)^{- {\ep \ov 2}} - {\m^\ep \ov 2 \pi \ep} 
\ \bgr] , \nn \\
G_1 (x,x') & = & v^\ah (x,x') \ \bgl[ \
{\G (-1 + {\ep \ov 2}) \ov 16 \pi^{D \ov 2}} 
\ \bl( 2 \s (x,x') \br)^{1 - {\ep \ov 2}} + 
{\m^\ep \ov 4 \pi \ep} \ \s (x,x') \ \bgr] .
\lbl{eqn:defG0G1}
\eea

Simple pole terms in $\ep$ are subtracted in the definitions of 
$G_0$ and $G_1$ so that they are finite in the limit $\ep \ra 0$.
These poles reflect the infrared singularities due to the large 
$\t$ behavior of the expansion (\ref{eqn:asyexp}). Since we have
subtracted the infrared singularities of the Green's functions here, 
we can safely identify 
the short distance singularities of the effective action 
as the poles in $\ep$ in the following sections. 
We have also introduced the dimensional regularization scale 
$\m$ to ensure the consistency of the dimension. We find the well-defined 
non-singular limits of (\ref{eqn:defG0G1}) as $\ep \ra 0$ to be 
\bea
G_0 & = & - {v^\ah \ov 4 \pi} \bl\{ \ 
\log (\m^2 \cdot 2 \s ) \ + \ \log \ \pi \ + \ \g \ \br\} \ , 
\nn \\ 
G_1 & = & {v^\ah \ov 16 \pi} \ 2 \s \bl\{ \ 
\log (\m^2 \cdot 2 \s ) \ + \ \log \ \pi \ - \ 1 \ + \ \g \ 
\br\} \ . 
\eea
The terms which contain logarithms cause the double-pole singularities 
in the two-loop calculations.

In the expansion (\ref{eqn:expG}), we have separated the potentially 
singular pieces in  the short distance limit $\s \ra 0$ from 
the non-singular ones. Only $G_0$ and $G_1$ produce the short 
distance singularities and they depend locally on the metric. 
On the other hand, the remainder $\ol G$ is regular for $x \sim x'$ 
even though it contains two derivatives.  It 
depends on the global aspects of the manifold in general.
We note that $\ol G$ is also constructed to be finite in the limit 
$\ep \ra 0$ as in the cases of $G_0$ and $G_1$.

Combining the definitions (\ref{eqn:defG0G1}) with (\ref{eqn:defd}), 
$G_0$ and $G_1$ are found to satisfy the followings:
\bea
(- \nab^2 \ + \ v^{-\ah} \nab^2 v^\ah ) \ G_0 & = & 
\dt^D , \nn \\
(\nab_\m \ - \ v^{-\ah} \nab_\m v^\ah ) \ G_1 & = & 
- \ah \ \s_\m G_0 , \nn \\
(- \nab^2 \ + \ v^{-\ah} \nab^2 v^\ah ) \ G_1 & = & 
G_0 \ - \ {\m^\ep \ov 4 \pi} \ v^\ah .
\lbl{eqn:rel1}
\eea
These relations are useful for us to derive the short distance 
singularities of the products of $G_0$'s and $G_1$'s with 
certain numbers of  
derivatives. These singular products of the functions $G_0 , G_1$ 
correspond to the short distance divergences which arise after the 
momentum integrations in multi-loop calculations. We 
perform the loop integrations in a coordinate space. These singular products 
form a basis of our calculation together with the coincidence limits. 
(For derivation, see appendix \ref{sec:sin}.)

By analytic continuation in $D$ from $D < 0$ and using the relations 
(\ref{eqn:rel1}), we find that for $G_0$,
\bea
G_0 | & \sim & - {\m^\ep \ov 2 \pi \ep} , \nn \\
\nab_\m G_0 | & \sim & 0 , \nn \\
\nab_\m \nab_\n G_0 | & \sim & {\m^\ep \ov 12 \pi \ep} \h R_\mn,
\lbl{eqn:limG0}
\eea
Similarly, the continuation from $D < 2$ shows that for $G_1$,
\bea
& & G_1 | \ \sim \ \nab_\m G_1 | \ \sim \ 0 , \nn \\
& & \nab_\m \nab_\n G_1 | \ \sim \ {\m^\ep \ov 4 \pi \ep} \h g_\mn .
\lbl{eqn:limG1}
\eea
Here we have used the coincidence limits of $\s$ and $v$ with 
the certain numbers of derivatives as shown in (\ref{eqn:clim1a})
and (\ref{eqn:clim1b}).

The coincidence limits (\ref{eqn:limG0}), (\ref{eqn:limG1}) of $G_0$ and 
$G_1$, together with the ones for $a_n$'s, easily lead to the limits 
of the Green's function.
\bea
G_\Dt | & \sim & - {\m^\ep \ov 2 \pi \ep} \ a^\Dt_0 | \ 
+ \ \ol G_\Dt | , \nn \\
\nab_\m G_\Dt | & \sim & \nab_\m \ol G_\Dt | , \nn \\
\nab_\m G_\Dt {\ola \nab}_{\n '} | & \sim & 
- {\m^\ep \ov 12 \pi \ep} \h R_{\m \n '} a^\Dt_0 |  
+ {\m^\ep \ov 2 \pi \ep} \ \nab_{\n '} \nab_\m a^\Dt_0 |  -  
{\m^\ep \ov 4 \pi \ep} \ \h g_{\m \n '} a^\Dt_1 |  +  
\nab_\m \ol G_\Dt {\ola \nab}_{\n '} | .
\lbl{eqn:limG}
\eea
Here the coincidence limits of the Seeley coefficients 
which can be obtained from (\ref{eqn:defSc1}) and (\ref{eqn:defSc2})
are given by 
%(For derivation, see appendix \ref{sec:sin}.)
\bea
a^\Dt_0 | & = & I \ , \nn \\ 
\nab_\m a^\Dt_0 | & = & 0 \ , \nn \\ 
\nab_\m \nab_\n a^\Dt_0 | & = & \ah \cR_\mn \ , \nn \\ 
a^\Dt_1 | & = & P \ - \ {1 \ov 6} \h R I \ .
\lbl{eqn:lima}
\eea
$\cR_\mn$ is defined through the commutation relation of the 
covariant derivatives for the fields on which the 
quadratic operator $\Dt_x$ acts:
\be
[\nab_\m , \nab_\n ] \ = \ \cR_\mn .
\lbl{eqn:com}
\ee
As is seen from the first relation of (\ref{eqn:limG}), 
the coincidence limit of $\ol G$ appears in
the finite non-local part of the one-loop effective action. 
It is also clear that the knowledge of the 
coincidence limits (\ref{eqn:limG}) of 
the Green's function suffices to obtain the
one-loop divergences.

%In addition, we can find similar relations for the non-local part 
%$\ol G$ from (\ref{eqn:defG0G1}), (\ref{eqn:rel1}), (\ref{eqn:defSc1}) 
%and (\ref{eqn:defSc2}).
%\bea
%\Dt_x \ol G_\Dt & = & - G_1 \ v^{- \ah} \ 
%\Dt_x (v^\ah a^\Dt_1) \ - \ {\m^\ep \ov 4 \pi} v^\ah a^\Dt_1 , \nn \\
%\Dt_x \ol G_\Dt | & = & {\m^\ep \ov 4 \pi} 
%\bl(- P + {1 \ov 6} \h R \ I \br) .
%\lbl{eqn:nl}
%\eea
%We note that $\ol G$ with the operator $\Dt_x$ acting on has a local 
%coincidence limit. 

In general, a multi-loop amplitude is the multiple integral of 
a multi-component function of some spacetime points. However the divergent 
part of the amplitude after the loop integrations is necessarily the one 
of a single-component function of a spacetime point. So the covariant 
Taylor expansion \cite{BV} around a resulting point is particularly useful, 
when we obtain the divergent part of the amplitude. This is because 
divergent parts depend only on the short distance behavior of the theory.  
The manifest covariance in the background fields greatly reduces 
the amount of the necessary calculations. 
%the covariant method 
%itself has the advantages that make our vision clear. 
The covariant Taylor expansion is a series which consists of the geodetic 
intervals and the expansion coefficients made of the background fields. 
The expansion coefficients are usually the field strength or the derivatives 
of it. 

In dimensional regularizations, there remain only logarithmic 
divergences. In the theories near two dimensions, logarithmically 
divergent terms are only those of the mass dimension two, which are made of 
the background fields. Therefore 
we only need to expand the Green's functions 
until the 
expansion coefficients have the mass dimension two
to obtain the divergent parts of the amplitude. 
These observations motivate us to apply
the covariant calculation method to the $2$-loop analysis 
of $(2+\ep)$-dimensional quantum gravity. 
In the remaining part of this section, 
we spell out necessary ingredients 
for such an investigation.

%\newpage
%\vspace{1cm}

\subsection{Applications to $(2 + \ep )$-dimensional Quantum Gravity}

Our starting point is the following action for gravitational fields, 
which is obtained from the pure Einstein action by reparametrizing 
the conformal mode $(\psi )$ so that its kinetic term becomes 
canonical\cite{AKKN}:
\be
I_{\rm gravity} \ = \ 
{\m^\ep \ov G} \idx \rgh \ 
\bgl\{ \til R (1 \ + \ a \ps \ + \ \ep b \ps^2 ) \ - \ 
\ah \til g^\mn \d_\m \ps \d_\n \ps \bgr\} .
\lbl{eqn:gra}
\ee
Classically parameters $a$ and $b$ take the values $a^2 \ = \ 4 \ep b \ = \ 
\ep /2(D-1)$ while $a^2$ receives the quantum correction at the one loop level.
We treat the conformal mode and the rest of the degrees of freedom
of the metric
separately and denote the latter as 
$\tilde{g}_{\mu\nu} = \hat{g}_{\mu\rho}({e^h})^{\rho}~_{\nu}$
where $h_{\mu\nu}$ is a tracelss symmetric tensor.
$\tilde{R}$ is the scalar curvature made out of $\tilde{g}_{\mu\nu}$.
We also couple $c$ copies of matter fields in the conformally 
invariant way as:
\be
I_{\rm matter} \ = \ 
{\m^\ep \ov G} \idx \rgh \ 
\bgl\{ \ah \til g^\mn \d_\m \vp_i \d_\n \vp_i \ - \ 
\ep b \vp_i^2 \til R \bgr\} ,
\lbl{eqn:ma}
\ee
The subscript $i$ runs from $1$ to $c$. We note that the matter 
fields decouple from the conformal mode.

For the two-loop calculations, we need to expand the action up to 
the quartic order in quantum fields. For this purpose, we exploit 
the following formula in our parameterization:
\bea
& & \tilde{R} \nn \\
& = & \h R - \nab_\m \nab_\n h^\mn - h^\m{}_\n \h R^\n{}_\m  \nn \\ 
& + & {1 \over 4 } \nab_\r h^\m{}_\n \nab^\r h^\n{}_\m 
- \ah \nab_\m h^\m{}_\r \nab_\n h^{\n \r} 
+ \ah \h R^\s{}_{\m \n \r} h^\r{}_\s h^\mn 
+\nab_\m (h^\m{}_\n \nab_\r h^\n{}^\r) \nn \\ 
& - & {1 \ov 4} h^\mn \nab_\m h^\r{}_\s \nab_\n h^\s{}_\r 
+ \ah h^\mn \nab_\m h^\r{}_\s \nab_\r h^\s{}_\n 
- {1 \ov 6} (h^3)^\m{}_\n \h R^\n{}_\m 
- {1 \ov 6} \nab_\m \nab_\n (h^3)^\mn \nn \\ 
& - & {1 \ov 6} (h^2)^\mn \nab_\n h^\r{}_\s \nab_\r h^\s{}_\m 
+ {1 \ov 12} \nab_\m h^\n{}_\r h^\r{}_\s \nab_\n h^\s{}_\k h^{\k \m} 
- {1 \ov 8} h^\m{}_\n \nab_\r h^\n{}_\s \nab_\m h^\s{}_\k h^{\k \r} 
\nn \\
& - & {1 \ov 24} \nab_\m h^\n{}_\r (h^2)^\r{}_\s \nab_\n h^{\s \m} 
+ {1 \ov 24} \nab_\m h^\n{}_\r \nab^\m h^\r{}_\s (h^2)^\s{}_\n 
- {1 \ov 24} \nab_\m h^\n{}_\r h^\r{}_\s \nab^\m h^\s{}_k h^\k{}_\n
\nn \\
& + & {1 \ov 8} (h^2)^\mn \nab_\m h^\r{}_\s \nab_\n h^\s{}_\r 
+ {1 \ov 24} (h^4)^\m{}_\n \h R^\n{}_\m 
+ {1 \ov 24} \nab_\m \nab_\n (h^4)^\mn 
+ \ O(h^5).
\lbl{eqn:expR}
\eea

In the background field method, the gauge fixing terms are chosen to 
fix the quantum field gauge invariance and to be covariant 
with respect to the background fields. 
In order to cancel the $h \ps$-mixing term with derivatives, we adopt a 
Feynmann-like gauge:
\bea
I_{\rm g.f.} & = & 
{\m^\ep \ov G} \idx \rgh \ 
\ah (\nab_\m h^\m{}_\r - a \d_\r \ps) 
(\nab_\n h^{\n \r} - a \d^\r \ps), \nn \\
& = & {\m^\ep \ov G} \idx \rgh \ 
\bgl\{ \ah \nab_\m h^\m{}_\r \nab_\n h^{\n \r} 
+ a \ps \nab_\m \nab_\n h^\mn  \nn \\
& & + \ah {\ep \ov 2(D-1)} \d_\m \ps \d^\m \ps 
- \ah {A G \ov 2(D-1)} \d_\m \ps \d^\m \ps \bgr\}. 
\lbl{eqn:gfa}
\eea
Here the last term is a finite one-loop term necessary to guarantee 
the general covariance up to the one-loop level. This follows 
from the result of ref.\cite{KKN3} where $a^2$ is related to 
the $\b $ function of the gravitational coupling constant $G$.
\bea
a^2 & = & {1 \ov 2(D-1)G} \ \m {d G \ov d \m } \ , \nn \\ 
& = & {1 \ov 2(D-1)} (\ep - A G + \lts ) \ .
\lbl{eqn:expa2}
\eea
The ``linear dilaton" like coupling $a$ is chosen to 
makes the conformal anomaly with respect to the background 
metric vanish. The parameter $A$ is determined to 
be ${25-c \ov 24 \pi}$ from the one-loop calculation.

Next let us consider the gauge 
transformation of the theory in order to obtain F.P. ghost terms.
The local gauge symmetry of gravity 
theories is the invariance under the general coordinate transformation. 
The metric changes under the infinitesimal shifts of the coordinates as
\be
\dt g_\mn = \d_\m \ep^\r g_{\r \n } + g_{\m \r } \d_\n \ep^\r
+ \ep^\r \d_\r g_\mn \ .
\ee
This leads to the gauge transformations 
of $\til g_\mn$, $\ps$ and $\vp_i $ fields as:
\bea
\dt \til g_\mn & = & 
\d_\m \ep^\r \til g_{\r \n } + \til g_{\m \r } \d_\n \ep^\r
+ \ep^\r \d_\r \til g_\mn - {2 \ov D} \nab_\r \ep^\r \til g_\mn , \nn \\
\dt \ps & = & 
\ep^\m \d_\m \ps + \{(D-1)a + {\ep \ov 4} \ps \} 
{2 \ov D} \nab_\m \ep^\m , \nn \\ 
\dt \vp_i & = & 
\ep^\m \d_\m \vp_i + ({\ep \ov 4} \vp_i ) 
{2 \ov D} \nab_\m \ep^\m .
\lbl{eqn:gtr1}
\eea
So we can derive the gauge transformation of  $h^\m{}_\n$ 
field from that of $\til g_\mn$:
\bea
\dt h^\m{}_\n & = & 
\nab^\m \ep_\n + \nab_\n \ep^\m + \ep^\r \nab_\r h^\m{}_\n 
- {2 \ov D} \dt^\m{}_\n \nab_\r \ep^\r \nn \\
& + & \ah [\nab \ep , h]^\m{}_\n \ 
+ \ \ah [h,(\nab \ep )^t]^\m{}_\n  \nn \\
& + & {1 \ov 12} [h,[h,\nab \ep ]]^\m{}_\n \ 
+ {1 \ov 12} [h,[h,(\nab \ep )^t]]^\m{}_\n \ 
+ \ O(h^3) .
\lbl{eqn:gtr2}
\eea

As a result, we find the Faddeev-Popov ghost action from the gauge fixing 
condition (\ref{eqn:gfa}) and the gauge transformations (\ref{eqn:gtr1}), 
(\ref{eqn:gtr2}) to be
\bea
I_{\rm gh.} & = & 
{\m^\ep \ov G} \idx \rgh \ 
\{ \bar \et_\m \nab^2 \et^\m \ - \ \bar \et^\m \h R_\mn \et^\n 
- \nab_\m \bar \et^\n \et^\r \nab_\r h^\m{}_\n  \nn \\ 
& - & \ah \nab_\m \bar \et^\n 
(\nab^\m \et^\r \ h_{\r \n } - h^{\m \r } \nab_\r \et_\n )
- \ah \nab_\m \bar \et^\n 
(h^\m{}_\r \nab_\n \et^\r - \nab_\r \et^\m \ h^\r{}_\n )  \nn \\
& - & {1 \ov 12} \nab_\m \bar \et^\n (h^2)^\m{}_\r \nab^\r \et_\n 
+ {1 \ov 6} \nab_\m \bar \et^\n h^\m{}_\r \nab^\r \et_\s h^\s{}_\n 
- {1 \ov 12} \nab_\m \bar \et^\n \nab^\m \et_\r (h^2)^\r{}_\n  \nn \\
& - & {1 \ov 12} \nab_\m \bar \et^\n (h^2)^\m{}_\r \nab_\n \et^\r 
+ {1 \ov 6} \nab_\m \bar \et^\n h^\m{}_\r \nab_\s \et^\r h^\s{}_\n 
- {1 \ov 12} \nab_\m \bar \et^\n \nab_\r \et^\m (h^2)^\r{}_\n  \nn \\
& + & a \nab_\m \bar \et^\m \ \et^\n \d_\n \ps 
+ {\ep a \ov 2D} \nab_\m \bar \et^\m \nab_\n \et^\n \ \ps 
- {A G \ov D} \nab_\m \bar \et^\m \nab_\n \et^\n \} \  .
\lbl{eqn:ga}
\eea
Here, we have used the expansion (\ref{eqn:expa2}) of $a^2$ and 
obtained a finite one-loop term as in the gauge-fixing terms 
(\ref{eqn:gfa}). 
%$a^{(0)}$ denotes the tree-level term of the 
%expansion (\ref{eqn:expa2}),{\it i.e.} $a^{(0)} = \sr{{\ep \ov 2(D-1)}}$.

We also need the expansion of the kinetic terms of $\ps$ and matter 
fields up to the quartic order. They are easily found, from 
(\ref{eqn:gra}) and (\ref{eqn:ma}), to be 
\bea
& & {\m^\ep \ov G} \idx \rgh 
\bgl\{ - \ah \d_\m \ps \d^\m \ps \ 
+ \ \ah \d_\m \vp_i \d^\m \vp_i  \nn \\ 
& & \qq \qq \qq + \ah h^\mn \d_\m \ps \d_\n \ps \ 
- \ \ah h^\mn \d_\m \vp_i \d_\n \vp_i  \nn \\ 
& & \qq \qq \qq - {1 \ov 4} (h^2)^\mn \d_\m \ps \d_\n \ps \ 
+ \ {1 \ov 4} (h^2)^\mn \d_\m \vp_i \d_\n \vp_i \bgr\} \ . 
\eea

Finally, we show the known one-loop counter terms we have determined in \cite{AKKN}
which consist of the gravitational coupling constant renormalization
and the linear wave function renormalization of the fields.
\bea
I_{\rm c.t.} & = & - {25 - c + 8 \ov 24 \pi \ep} \idx \rgh 
\bgl( {1 \ov 4} \nab_\r h^\m{}_\n \nab^\r h^\n{}_\m 
- \ah \nab_\m h^\m{}_\r \nab_\n h^{\n \r} 
+ \ah \h R^\s{}_{\m \n \r} h^\mn h^\r{}_\s \bgr) \nn \\
& & + {1 \ov 6 \pi \ep } \idx \rgh \ 
a \ps (\nab_\m \nab_\n h^\mn + h^\m{}_\n \h R^\n{}_\m)  \nn \\
& & - {1 \ov 12 \pi \ep } \idx \rgh \ 
\bl( \nab_\m \bar \et_\n \nab^\m \et^\n \ 
+ \ \bar \et^\m \h R_\mn \et^\n \br) \ .
\lbl{eqn:cta}
\eea
There is no renormalization of the gauge fixing term (\ref{eqn:gfa})
which is as expected in generic gauge theories.

%\vspace{1cm}

Now, based on the above expansions in quantum fields, we will 
apply the heat kernel method to the two-loop renormalizations 
in $(2+\ep )$-dimensional quantum gravity.

First, we collect the quadratic terms in quantum fields to 
define exact propagators:
\bea
%(I)_2 & = & 
& & (I_{\rm gravity} + I_{\rm matter} 
+ I_{\rm g.f.} + I_{\rm gh.})_2 ,  \nn \\ 
& = & {\m^\ep \ov G} \idx \rgh \ 
\bgl[ - {1 \ov 4} h_\mn (I^{\mn ,\rs } \nab^2 
+ 2 I^\mn{}_{, \ab} \h R^\a{}_\g{}^\b{}_\dt I^{\cd ,\rs } ) h_\rs 
- a \ps h_\mn I^\mn{}_{, \rs} \h R^\rs  \nn \\
& & + \ah \ps \bgl\{ \bgl(1 - a^2 \bgr) \nab^2 \ 
+ \ 2 \ep b \h R \bgr\} \ps  \nn \\ 
& & + \bar \et^\m (\h g_\mn \nab^2 \ - \ \h R_\mn ) \et^\n  \nn \\
& & - \ah \vp_i (\dt_{ij} \nab^2 \ + \ 2 \ep b \h R \dt_{ij} ) \vp_j 
\bgr] \ ,
\eea
where 
\be
I^\mn{}_{,\rs } = \ah \dt^\m{}_\r \dt^\n{}_\s 
+ \ah \dt^\m{}_\s \dt^\n{}_\r - {1 \ov D} \h g^\mn \h g_\rs
\ee 
is the identity for the traceless symmetric tensors 
in a $D$-dimensional curved space. 

For later convenience, we rescale the fields as 
\bea
& & h_\mn \ra \sr{2 G} h_\mn \ \ , \ \ 
\ps \ra i \bl(1 + {a^2 \ov 2} \br) \sr{G} \ps , \nn \\ 
& & \bar \et^\m , \et^\m \ra 
\sr{G} \bar \et^\m , \sr{G} \et^\m \ \ , \ \ 
\vp_i \ra \sr{G} \vp_i \ .
\eea
So we can define the exact propagators 
$G_{\mn ,\r ' \s '} (x,x') , \ G^{\rm gh}_{\m \n '} (x,x') , \ 
G_{\ps \ps '} (x,x') , \ G_{i j'} (x,x')$ for $h_\mn$, ghost, $\ps$ and 
$\vp_i$ fields respectively, 
up to $O(\ep )$, which is enough to obtain divergent contributions.
\bea
\Dt_{AC} G^C{}_{B'} (x,x') & = & - I_{A B'} \ \dt^D (x,x') \ , \nn \\ 
\Dt^{\rm gh}_{\m \r} \ G^{{\rm gh} \ \r}{}_{\n '} (x,x') & = & 
- \h g_{\m \n '} \ \dt^D (x,x') \ , \nn \\ 
\Dt_{ik} G_{kj'} (x,x') & = & - \dt_{ij'} \ \dt^D (x,x') \ .
\eea
Here, the quadratic operators $\Dt$ are as follows.
\bea
\Dt_{AB} & = & 
{\left(
\begin{array}{cc}
\Dt_{\mn , \rs } & \Dt_{\mn , \ps } \\ 
\Dt_{\ps , \rs } & \Dt_{\ps \ps } 
\end{array}
\right)} \ , \nn \\ 
& = & 
{\left(
\begin{array}{cc}
I_{\mn , \rs } & 0 \\ 
0 & 1 
\end{array}
\right)} 
\nab^2 \ + \ 
{\left(
\begin{array}{cc}
2 I_{\mn , \ab} \h R^\a{}_\g{}^\b{}_\dt I^\cd{}_{, \rs } & 
i \sr{2} \bl(1 + {a^2 \ov 2} \br) a I_{\mn ,}{}^\rs \h R_\rs \\ 
i \sr{2} \bl(1 + {a^2 \ov 2} \br) a I_{\rs ,}{}^\mn \h R_\mn & 
2 \ep b \h R 
\end{array}
\right)} \ , \nn \\ 
& \equiv & 
I_{AB} \nab^2 \ + \ P_{AB} (x) \ , \nn \\ 
G_{A B'} (x,x') & = & 
\left(
\begin{array}{cc}
G_{\mn , \r ' \s '} (x,x') & G_{\mn , \ps '} (x,x') \\ 
G_{\ps , \r ' \s '} (x,x') & G_{\ps \ps '} (x,x') 
\end{array}
\right) \ , \nn \\ 
\Dt^{\rm gh}_\mn & = & 
\h g_\mn \nab^2 \ - \ \h R_\mn \ , \nn \\ 
& \equiv & 
\h g_\mn \nab^2 \ + \ P_\mn \ , \nn \\ 
\Dt_{ij} & = & 
\dt_{ij} \nab^2 \ + \ 2 \ep b \h R \dt_{ij} \ , \nn \\ 
& \equiv & 
\dt_{ij} \nab^2 \ + \ P_{ij} \ .
\lbl{eqn:qop}
\eea
We note the traceless property of the quadratic operator 
$\Dt_{\mn , \rs }$ and the exact propagator $G_{\mn ,\r ' \s '} (x,x')$ for 
$h_\mn$ field.
\bea
& & \h g^\mn (x) \ \Dt_{\mn , \rs } \ 
= \ \h g^\rs (x) \ \Dt_{\mn , \rs } \ = \ 0 \ , \nn \\ 
& & \h g^\mn (x) \ G_{\mn ,\r ' \s '} (x,x') \ 
= \ \h g^{\r ' \s '} (x') \ G_{\mn ,\r ' \s '} (x,x') \ = \ 0 \ . 
\eea
These constraints originate from the traceless property of $h_\mn$ field. 
\bea
& & \h g^\mn (x) \ h_\mn (x) \ 
= \ \h g^{\r ' \s '} (x') h_{\r ' \s '} (x') \ = \ 0 \ , \nn \\ 
& & G_{\mn ,\r ' \s '} (x,x') \ 
= \ <h_\mn (x) h_{\r ' \s '} (x')> \ . 
\eea

From the expansion (\ref{eqn:expG}), we should also note the traceless 
property of the nonlocal part of $h_\mn$ field propagator. 
\be
\h g^\mn (x) \ \ol G_{\mn ,\r ' \s '} (x,x') \ 
= \ \h g^{\r ' \s '} (x') \ \ol G_{\mn ,\r ' \s '} (x,x') \ = \ 0 \ . 
\ee
It is clear that we can obtain similar relations for $h_\mn$-$\ps$ 
mixing parts.
\bea
& & \h g^\mn (x) \ \Dt_{\mn , \ps } \ 
= \ \h g^\rs (x) \ \Dt_{\ps , \rs } \ = \ 0 \ , \nn \\ 
& & \h g^\mn (x) \ G_{\mn ,\ps '} (x,x') \ 
= \ \h g^{\r ' \s '} (x') \ G_{\ps ,\r ' \s '} (x,x') \ = \ 0 \ , \nn \\ 
& & \h g^\mn (x) \ \ol G_{\mn ,\ps '} (x,x') \ 
= \ \h g^{\r ' \s '} (x') \ \ol G_{\ps ,\r ' \s '} (x,x') \ = \ 0 \ . 
\eea

%\vspace{1cm}

Next, let us consider the Seeley coefficients. As is seen in the 
previous section, the $0$th Seeley coefficient $a^\Dt _0$ is the 
geodetic parallel displacement matrix. Solving the definition 
(\ref{eqn:defSc1}), we obtain the $0$th Seeley coefficients for the 
fields as 
\bea
a_{0 \ \mn , \r ' \s '} & = & 
\ah \g_{\m \r '} \g_{\n \s '} 
+ \ah \g_{\m \s '} \g_{\n \r '} 
- {1 \ov D} \h g_\mn (x) \h g_{\r ' \s '} (x') \ , \nn \\ 
a_{0 \ \mn , \ps '} & = & a_{0 \ \ps , \r ' \s '} \ = \ 0 \ , \nn \\ 
a_{0 \ \ps \ps '} & = & 1 \ , \nn \\ 
a^{\rm gh}_{0 \ \m \n '} & = & \g_{\m \n '} \ , \nn \\ 
a_{0 \ ij'} & = & \dt_{ij'} \ . 
\eea
Here $\h g_\mn$ is the background metric, and $\g_{\m \n '}$ is the 
{\it geodetic parallel displacement bi-vector} defined by the differential 
equation:
\bea
\s^\r \nab_\r \g_{\m \n '} & = & 0 \ , \nn \\ 
\lim_{x' \ra x} \g_{\m \n '} & = & \h g_{\m \n '} (x) \ . 
\lbl{eqn:defp}
\eea
The bi-vector $\g_{\m \n '}$ effects the parallel displacement of 
a contravariant vector $A^{\n '}$ at $x'$ to the covariant form at $x$ 
along the geodesic from $x'$ to $x$. 

Again, we note that the coefficient $a_{0 \ \mn , \r ' \s '}$ for 
$h_\mn$ field is traceless with respect to the background metric.
\be
\h g^\mn (x) \ a_{0 \ \mn , \r ' \s '} (x,x') \ 
= \ \h g^{\r ' \s '} (x') \ a_{0 \ \mn , \r ' \s '} (x,x') \ = \ 0 \ . 
\ee

%\vspace{1cm}

Finally, we can find the coincidence limits of various Seeley 
coefficients from (\ref{eqn:qop}). They are the most important 
ingredients of our two-loop calculations, together with the limits 
of $\s , v$ and the singular products of the functions $G_0 , G_1$. 

\underline{$h_\mn$ field}
\bea
a_{0 \ \ab , \g ' \dt '} | & = & I_{\ab , \g ' \dt '} \ , \nn \\ 
\nab_\m \nab_\n a_{0 \ \ab , \g ' \dt '} | 
& = & \ah (\cR_\mn )_{\ab , \g ' \dt '} \ , \nn \\ 
& = & {1 \ov 4} 
(\h R_{\mn \a \g '} \h g_{\b \dt '} 
+ \h R_{\mn \a \dt '} \h g_{\b \g '} 
+ \h R_{\mn \b \g '} \h g_{\a \dt '} 
+ \h R_{\mn \b \dt '} \h g_{\a \g '} ) 
\ , \nn \\ 
a_{1 \ \ab , \g ' \dt '} | & = & 
P_{\ab , \g ' \dt '} - {1 \ov 6} \h R \ I_{\ab , \g ' \dt '} \ .
\eea

\underline{$h_\mn$-$\ps$ fields}
\be
a_{1 \ \ab , \ps '} | = P_{\ab , \ps '} \ . 
\lbl{eqn:climhpsi}
\ee

\underline{$\ps$ field}
\bea
a_{0 \ \ps \ps '} | & = & 1 \ , \nn \\ 
\nab_\m \nab_\n a_{0 \ \ps \ps '} | & = & 0 \ , \nn \\ 
a_{1 \ \ps \ps '} | & = & P_{\ps \ps '} - {1 \ov 6} \h R \ . 
\eea

\underline{ghost field}
\bea
a_{0 \ \a \b '} | 
& = & \h g_{\a \b '} \ , \nn \\ 
\nab_\m \nab_\n a_{0 \ \a \b '} | 
& = & \ah (\cR_\mn )_{\a \b '} \ , \nn \\ 
& = & \ah \h R_{\a \b ' \mn } \ , \nn \\ 
a_{1 \ \a \b '} | & = & 
P_{\a \b '} - {1 \ov 6} \h R \ \h g_{\a \b '} \ . 
\eea

\underline{$\vp_i$ fields}
\bea
a_{0 \ i j'} | & = & \dt_{i j'} \ , \nn \\ 
\nab_\m \nab_\n a_{0 \ i j'} | & = & 0 \ , \nn \\ 
a_{1 \ i j'} | & = & P_{i j'}- {1 \ov 6} \h R \ \dt_{i j'} \ . 
\eea
Here, we note that $\nab_\m a_0 | = 0$ for all fields. This is easily 
seen from the fact that no local and covariant quantity has the mass 
dimension one.

%\vspace{3cm}

%\newpage

\section{Two-loop Renormalization of Pure Einstein Gravity}
\setcounter{equation}{0}
\lbl{sec:ren}

In this section, we perform the two-loop renormalization of the 
pure Einstein gravity theory (without the cosmological constant).
In order to do so, we use the heat 
kernel method discussed in the previous section. 
In the heat kernel method, the evaluations of the non-local divergences 
of the effective action at the two loop level
are exactly the same as the ones needed 
for the one-loop quantum field renormalizations.
The nonlocal divergences will be shown to be 
cancelled by introducing the counter terms
we have already identified at the one loop level except the type
which is proportional to the field equation. 
We then shaw that such a non-local divergence can be
canceled by a non-linear wave function renormalization of $h_\mn$ field. 

Our results in this section hence explicitly show that the two-loop divergences 
of the pure Einstein gravity theory
can be made to be local by the 
one-loop quantum field renormalizations and the renormalization of the 
gravitational coupling constant.
The two loop level divergence of the effective action
is shown to be the Einstein action form and
hence can be canceled by the renormalization of the gravitational
coupling constant at the two loop level.

%\vspace{1cm}

\subsection{Two-loop Diagrams with $\ps$ or Matter Fields' Propagators}

\begin{figure}
 \leavevmode
 \epsfxsize = 10 cm
 \centerline{ \epsfbox{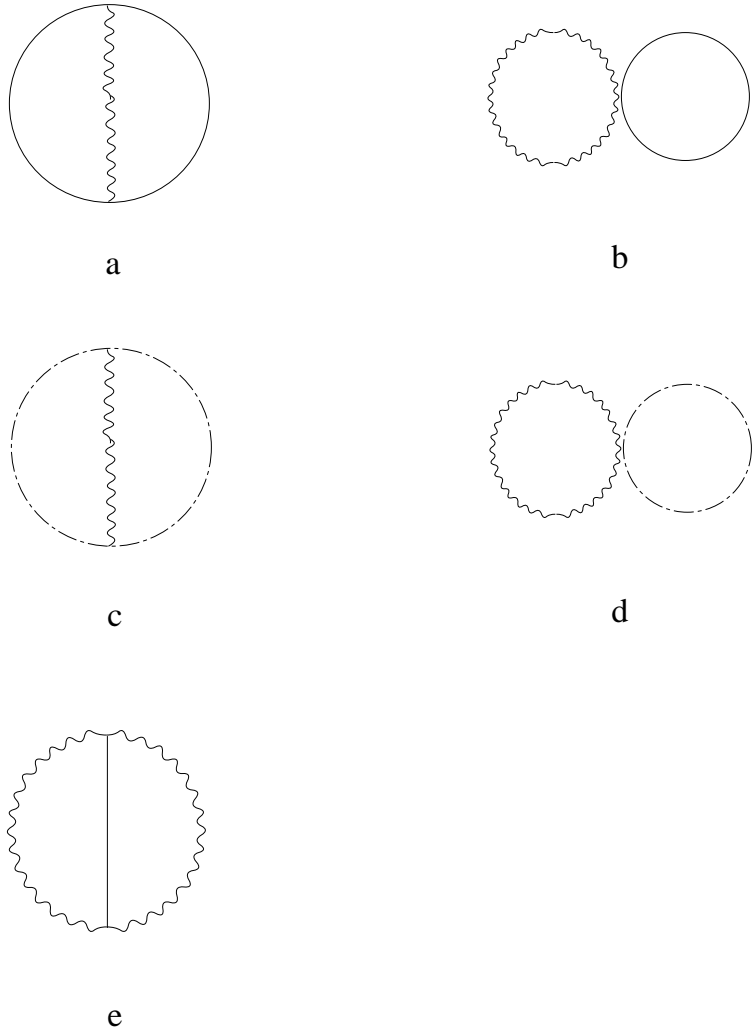} }
 \caption{Two-loop diagrams with $\ps$ or matter fields' propagators}
 \label{fig:m}
\end{figure}

First, we evaluate the two-loop diagrams with $\ps$ or matter fields' 
propagators. In fig.\ref{fig:m}, the wavy and solid lines denote the propagators
of $h_\mn $ and $\ps $ fields, 
while the dashed-and-dotted lines denote the propagators of the matter fields.
In our previous work, we evaluated the diagrams of 
figs.\ref{fig:m}{\rm a}-\ref{fig:m}{\rm d}, 
expanding the background metric around the flat one \cite{AKNT}. 
%After a large amount of calculation, we have obtained the results 
%of a renormalizable form. 
%Here we can perform the evaluations more easily and recognize 
%renormalization procedures explicitly. 
%
%\vspace{0.6cm}
Here we illustrate the evaluation of the amplitudes for 
fig.\ref{fig:m}{\rm a} in the heat kernel method. 

The amplitudes are 
given by the following three types of the multiple integrals:
\bea
& & \int d V \ d V'  \ 
<\d_{\m } \ps (x) \d_{\m '} \ps (x')> 
<\d_{\n } \ps (x) \d_{\n '} \ps (x')> 
<h_{\a \b } (x) h_{\a ' \b '} (x')>  \nn \\
& = & 
\int d V  \ d V' \  
\d_\m G_{\ps \ps '} {\ola \d}_{\m '} \ct 
\d_\n G_{\ps \ps '} {\ola \d}_{\n '} \ct 
G_{\a \b ,\a ' \b '} \ , \lbl{eqn:ex1} \\ 
& & \int d V  \ d V'  \ 
<\d_{\m } \ps (x) \d_{\m '} \ps (x')> 
<\d_{\n } \ps (x) \ps (x')> 
<h_{\a \b } (x) \nab_{\n '} h_{\a ' \b '} (x')>  \nn \\
& = & 
\int d V  \ d V' \ 
\d_\m G_{\ps \ps '} {\ola \d}_{\m '} \ct 
\d_\n G_{\ps \ps '} \ct 
G_{\a \b ,\a ' \b '} {\ola \nab}_{\n '} \ , \lbl{eqn:ex2} \\ 
& & \int d V  \ d V'  \ 
<\d_{\m } \ps (x) \ps (x')> 
<\d_{\n } \ps (x) \ps (x')> 
<h_{\a \b } (x) h_{\a ' \b '} (x')> \h R_{\r ' \s '} (x') \nn \\
& = & 
\int d V  \ d V' \ 
\d_\m G_{\ps \ps '} \ct 
\d_\n G_{\ps \ps '} \ct 
G_{\a \b ,\a ' \b '} \h R_{\r ' \s '} \ , \lbl{eqn:ex3}
\eea
where $d V$ denotes an invariant volume element $d^D x \sr{\h g}$. 

We can evaluate the divergent parts of the above three types of 
amplitudes for general subscripts. We first substitute the short distance 
expansion (\ref{eqn:expG}) of the exact propagators into the 
equations (\ref{eqn:ex1})-(\ref{eqn:ex3}). The expansions of them are found 
to consist of the products of the functions $G_0 , G_1$ 
and the Seeley coefficients and $\ol G$. 
In order to extract divergent contributions, let us look at the products 
of the Seeley coefficients and $\ol G$. We only have to keep the products of 
the quantities with the mass dimensions less than or equal to two, 
when we make the covariant Taylor expansions of them. This is seen 
from a simple power counting argument. 

Furthermore, we have to extract the singular behavior proportional to 
a delta function from the products of the functions $G_0 , G_1$. 
(See appendix \ref{sec:sin}.) The singular behavior with a delta 
function corresponds to the short distance singularity due to the
loop integrations. 
In this way, we obtain the products of the necessary Seeley coefficients and 
a delta function as the short distance behavior of the integrands of the multiple 
integrals. The delta functions with no derivative make it possible 
for us to take simple coincidence limits of the Seeley coefficients and $\ol G$. 
But in the case of the delta function with the derivatives, we have to pay 
a little more attention to take coincidence limits. For instance, we may use 
an identity as 
\be
f \nab^2 \dt^D \ = \ 
\nab^2 (f| \ct \dt^D ) - 2 \nab^\m (\nab_\m f| \ct \dt^D ) 
+ \nab^2 f | \ct \dt^D \ . 
\ee
In such a procedure, we can obtain the singular parts of the 
amplitudes (\ref{eqn:ex1})-(\ref{eqn:ex3}). 

As an example, we can write down the explicit form of the singular 
contributions of the amplitude (\ref{eqn:ex1}) as follows.
\bea
& & \int d^D V \ d^D V'  \ 
<\d_{\m } \ps (x) \d_{\m '} \ps (x')> 
<\d_{\n } \ps (x) \d_{\n '} \ps (x')> 
<h_{\a \b } (x) h_{\a ' \b '} (x')>  \nn \\
& \sim & 
\int d^D V  \ d^D V' \ \bgl[
(\dt^D {\rm part \ of \ } \g_\r{}^{ \r '} \d_\m G_0 {\ola \d}_{\r '} \ct 
\g_\s{}^{ \s '} \d_\n G_0 {\ola \d}_{\s '} \ct G_0 ) \ 
\h g^\r{}_{\m '} \h g^\s{}_{\n '} I_{\a \b , \a ' \b ' } \nn \\ 
& + &   
(\dt^D {\rm part \ of \ } \g_\r{}^{ \r '} \d_\m G_0 {\ola \d}_{\r '} \ct 
\g_\s{}^{ \s '} \d_\n G_0 {\ola \d}_{\s '} \ct G_1 ) \ 
\h g^\r{}_{\m '} \h g^\s{}_{\n '} P_{\a \b , \a ' \b ' } \nn \\ 
& + &   
(\dt^D {\rm part \ of \ } \g_\r{}^{ \r '} \d_\m G_0 {\ola \d}_{\r '} \ct 
\g_\s{}^{ \s '} \d_\n G_1 {\ola \d}_{\s '} \ct G_0 ) \ 
\h g^\r{}_{\m '} \h g^\s{}_{\n '} P_{\ps \ps } I_{\a \b , \a ' \b ' } \nn \\ 
& + &   
(\dt^D {\rm part \ of \ } \g_\r{}^{ \r '} \d_\n G_0 {\ola \d}_{\r '} \ct 
\g_\s{}^{ \s '} \d_\m G_1 {\ola \d}_{\s '} \ct G_0 ) \ 
\h g^\r{}_{\n '} \h g^\s{}_{\m '} P_{\ps \ps } I_{\a \b , \a ' \b ' } \nn \\ 
& + & 
(\nab_\m \nab_\n \dt^D {\rm \ part \ of \ } 
\g_\r{}^{ \r '} \d_\m G_0 {\ola \d}_{\r '} \ct 
\g_\s{}^{ \s '} \d_\n G_0 {\ola \d}_{\s '} ) \nn \\ 
& & {\rm with \ } \nab_\m \nab_\n \dt^D 
\longrightarrow 
\dt^D \ct \h g^\r{}_{\m '} \h g^\s{}_{\n '} 
(\nab_\m \nab_\n {\ol G}_{\a \b , \a ' \b ' } | + 
\nab_\n \nab_\m {\ol G}_{\a \b , \a ' \b ' } |) /2 
\nn \\ 
& + & 
(\dt^D {\rm part \ of \ } \g_\r{}^{ \r '} \d_\m G_0 {\ola \d}_{\r '} \ct 
\g_\s{}^{ \s '} \d_\n G_0 {\ola \d}_{\s '} ) \ 
\h g^\r{}_{\m '} \h g^\s{}_{\n '} {\ol G}_{\a \b , \a ' \b ' } | \nn \\ 
& + &   
(\dt^D {\rm part \ of \ } \g_\r{}^{ \r '} \d_\m G_0 {\ola \d}_{\r '} \ct 
\g_\s{}^{ \s '} \d_\n G_1 {\ola \d}_{\s '} ) \ 
\h g^\r{}_{\m '} \h g^\s{}_{\n '} P_{\ps \ps } {\ol G}_{\a \b , \a ' \b ' } | 
\nn \\ 
& + &   
(\dt^D {\rm part \ of \ } \g_\r{}^{ \r '} \d_\n G_0 {\ola \d}_{\r '} \ct 
\g_\s{}^{ \s '} \d_\m G_1 {\ola \d}_{\s '} ) \ 
\h g^\r{}_{\n '} \h g^\s{}_{\m '} P_{\ps \ps } {\ol G}_{\a \b , \a ' \b ' } | 
\nn \\ 
& - &   
(\dt^D {\rm part \ of \ } \g_\r{}^{ \r '} \d_\m G_0 {\ola \d}_{\r '} \ct 
G_0 ) \ 
\h g^\r{}_{\m '} I_{\a \b , \a ' \b ' } \nab_{\n '} \nab_\n {\ol G}_{\ps \ps } | 
\nn \\ 
& - &   
(\dt^D {\rm part \ of \ } \g_\r{}^{ \r '} \d_\n G_0 {\ola \d}_{\r '} \ct 
G_0 ) \ 
\h g^\r{}_{\n '} I_{\a \b , \a ' \b ' } \nab_{\m '} \nab_\m {\ol G}_{\ps \ps } | \ 
\bgr] \ . 
\lbl{eqn:expex1}
\eea
Here, the $\dt^D$ part denotes the singular one of the product of 
the functions $G_0 , G_1$, which is proportional to a delta function. Also, 
a quadratic derivative part of a delta function is meant by 
the $\nab_\m \nab_\n \dt^D$ part.

%\vspace{1cm}

As a result, the total divergent amplitude for fig.\ref{fig:m}{\rm a} is found to be 
\bea
& & {G \ov (4 \pi )^2} \idx \rgh \ {-7 + 96 b \ov 24 \ep } \h R 
\nn \\ 
& + & {G \ov 4 \pi } \idx \rgh \ \bgl(
- {1 \ov 12 \ep } \nab^2 \ol G_{\mn ,}{}^\mn | 
+ {1 \ov 6 \ep } \nab_\m \nab_\n \ol G{^\n{}_{\r ,}{}^{\m \r }} | \nn \\
& &\qq \qq \qq \ \ 
-{1 \ov 6 \ep } \h R_{\m \r \n \s } \ol G{^{\mn , \rs }} | 
- {1 \ov 6 \ep } \h R_\mn \ol G{^\m{}_{\r ,}{}^{\n \r }} | 
+ {1 \ov 12 \ep } \h R \ol G_{\mn ,}{}^\mn |  \nn \\
& &\qq \qq \qq \ \ 
+ {1 \ov \ep } \nab^2 \ol G_{\ps \ps } | 
\bgr) \ . 
\lbl{eqn:psi1}
\eea
Here, we note that there arise only the single-pole divergences. 
In general, the two-loop quantum corrections give rise to the double-pole divergences 
as well as the single-pole ones. However, the cancellation of the 
double-pole divergences is necessary to ensure the consistency of the 
theory, such as the finiteness of the $\b$ functions of the renormalization 
group. In the remaining part of this section, we will also find each type of 
the diagrams to yield only the single-pole divergences. 

%\vspace{1cm}
Let us proceed to the evaluation of the amplitudes 
of fig.\ref{fig:m}{\rm b}. They are of the form of the square of the one-loop 
amplitudes. Here, we also have the three types of the expectation values to consider.
\bea
& & \idx \rgh \ <\d_{\m } \ps (x) \d_{\n } \ps (x)> 
<h_{\a \b } (x) h_{\g \dt } (x)>  \nn \\
& = & 
\idx \rgh \ 
\d_\m G_{\ps \ps } {\ola \d}_{\n } | \ct 
G_{\a \b , \g \dt } | \ , \lbl{eqn:ex4} \\ 
& & \idx \rgh \ < \nab_{\m } h_{\a \b } (x) \nab_{\n } h_{\g \dt } (x)> 
< \ps (x) \ps (x)>  \nn \\
& = & 
\idx \rgh \ 
\nab_\m G_{ \a \b, \g \dt } {\ola \nab}_{\n } | \ct 
G_{\ps \ps } | \ , \lbl{eqn:ex5} \\ 
& & \idx \rgh \ <h_{\a \b } (x) h_{\g \dt } (x)> 
< \ps (x) \ps (x) > \h R_{\mn \rs } (x) \nn \\
& = & 
\idx \rgh \ 
G_{\ab ,\g \dt } | \ct 
G_{\ps \ps } | \ct \h R_{\mn \rs } \ . \lbl{eqn:ex6}
\eea
The evaluation of the above is much easier than that of fig.\ref{fig:m}{\rm a}. 
It is straightforward to perform it by using 
the coincidence limits (\ref{eqn:limG}) of the exact 
propagators.

As an example, we explicitly write down the divergent contributions 
of the amplitude (\ref{eqn:ex4}).
\bea
& & \idx \rgh \ <\d_{\m } \ps (x) \d_{\n } \ps (x)> 
<h_{\a \b } (x) h_{\g \dt } (x)>  \nn \\
& \sim & 
\idx \rgh \ \bgl[ {\m^{2 \ep} \ov (4 \pi )^2} 
{2 \ov 3 \ep^2} (\h R_\mn 
+ 3 \h g_\mn P_{\ps \ps } ) \ I_{\ab , \cd } \nn \\ 
& & \qq \qq \  
- {\m^\ep \ov 4 \pi } {1 \ov 3 \ep } (\h R_\mn + 3 \h g_\mn P_{\ps \ps }) 
\ {\ol G}_{\ab , \cd } | \nn \\ 
& & \qq \qq \  
+ {\m^\ep \ov 4 \pi } {2 \ov \ep } \nab_\n \nab_\m {\ol G}_{\ps \ps } | 
\ct I_{\ab , \cd } \bgr] \ . 
\lbl{eqn:expex4}
\eea
We have obtained the total divergent amplitude for fig.\ref{fig:m}{\rm b} as 
follows.
\bea
& & {G \ov (4 \pi )^2} \idx \rgh \ {1 - 24 b \ov 6 \ep } \h R \nn \\ 
& + & {G \ov 4 \pi } \idx \rgh \ \bgl(
{1 \ov 6 \ep } \h R_\mn \ol G{^\m{}_{\r ,}{}^{\n \r }} | 
- {1 \ov 12 \ep } \h R \ol G_{\mn ,}{}^\mn | 
- {1 \ov \ep } \nab^2 \ol G_{\ps \ps } | \bgr) \ . 
\lbl{eqn:psi2}
\eea

%\vspace{1cm}

We are left to consider the amplitudes for 
figs.\ref{fig:m}{\rm c},\ref{fig:m}{\rm d}. There is little difference 
between $\psi$ and matter fields except for the term: $a \psi \til R$. 
(See the actions (\ref{eqn:gra}) and (\ref{eqn:ma}).) It is irrelevant 
to the evaluation of these amplitudes, since this term is linear in $\psi$. 
As a result, we have found that the amplitudes for 
figs.\ref{fig:m}{\rm c},\ref{fig:m}{\rm d} are 
given by (\ref{eqn:psi1}) and (\ref{eqn:psi2}), respectively, 
multiplied by the number of matter fields $c$ and with 
$\ol G_{\ps \ps }$ replaced by $\ol G_{\vp \vp }$.

The amplitudes for fig.\ref{fig:m}{\rm e} give neither double-pole nor 
non-local divergence. This is because the products of the 
three-point vertices are proportional to $a^2 \simeq \ep /2 $.
Although $a^2$ receives the quantum correction, it can be ignored here
since it is at higher orders in $G$. 
In this way we have found the divergent amplitude of this diagram to be: 
\be
{G \ov (4 \pi )^2} \idx \rgh \ \bgl(-{1 \ov 2 \ep } \bgr) \h R \ . 
\lbl{eqn:psi3}
\ee
Here, we note that this result is obtained from the double-pole 
divergence suppressed by $a^2 \simeq \ep /2 $. Such underlying double-pole 
singularities arise only in (\ref{eqn:psi3}) and (\ref{eqn:g3}). 

%\vspace{1cm}

\subsection{Two-loop Diagrams with Only $h_\mn$ Field Propagators}

Secondly, we consider the two-loop diagrams made only of 
the propagators of $h_\mn$ field. 
\begin{figure}
 \leavevmode
 \epsfxsize = 9 cm
 \centerline{ \epsfbox{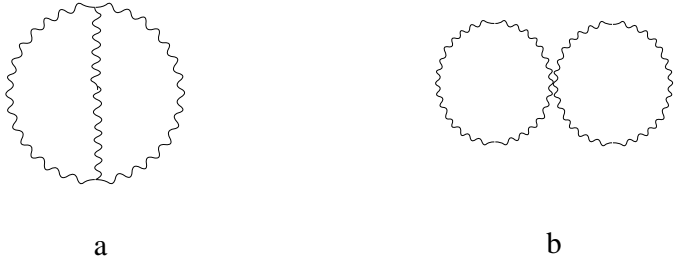}}
 \caption{Two-loop diagrams made only of $h_\mn$ field propagators}
 \label{fig:h}
\end{figure}

%In a quite similar way to the previous section, we can evaluate 
%the amplitudes of fig.\ref{fig:h}{\rm a}. 
%We only have to consider three types of expectation values.
%\bea
%& & \int d^D V \ d^D V'  \ 
%<\nab_{\m } h_{\a \b } (x) \nab_{\m '} h_{\a ' \b '} (x')> 
%<\nab_{\n } h_{\g \dt } (x) \nab_{\n '} h_{\g ' \dt '} (x')> 
%<h_{\k \l } (x) h_{\k ' \l '} (x')>  \nn \\
%& = & 
%\int d^D V  \ d^D V' \  
%\nab_\m G_{\ab ,\a ' \b '} {\ola \nab}_{\m '} \ct 
%\nab_\n G_{\cd ,\g ' \dt '} {\ola \nab}_{\n '} \ct 
%G_{\k \l ,\k ' \l '} \ , \nn \\ 
%& & \int d^D V  \ d^D V'  \ 
%<\nab_{\m } h_{\a \b } (x) \nab_{\m '} h_{\a ' \b '} (x')> 
%<\nab_{\n } h_{\g \dt } (x) h_{\g ' \dt '} (x')> 
%<h_{\k \l } (x) \nab_{\n '} h_{\k ' \l '} (x')>  \nn \\
%& = & 
%\int d^D V  \ d^D V' \ 
%\nab_\m G_{\ab ,\a ' \b '} {\ola \nab}_{\m '} \ct 
%\nab_\n G_{\cd ,\g ' \dt '} \ct 
%G_{\k \l ,\k ' \l '} {\ola \nab}_{\n '} \ , \nn \\ 
%& & \int d^D V  \ d^D V'  \ 
%<\nab_{\m } h_{\a \b } (x) h_{\a ' \b '} (x')> 
%<\nab_{\n } h_{\g \dt } (x) h_{\g ' \dt '} (x')> 
%<h_{\k \l } (x) h_{\k ' \l '} (x')> \h R_{\r ' \s '} (x') \nn \\
%& = & 
%\int d^D V  \ d^D V' \ 
%\nab_\m G_{\ab ,\a ' \b '} \ct 
%\nab_\n G_{\cd ,\g ' \dt '} \ct 
%G_{\k \l ,\k ' \l '} \h R_{\r ' \s '} \ . 
%\eea
%We can also evaluate the divergent parts of the above three types of 
%amplitudes for general subscripts. 

In a quite similar way to the previous subsection, we can evaluate 
the amplitudes for fig.\ref{fig:h}{\rm a}. We only have to consider three 
types of the expectation values. We can also evaluate the divergent parts of them for 
general subscripts. Using them, we have found the total divergent amplitude 
for fig.\ref{fig:h}{\rm a} to be 
\bea
& & {G \ov (4 \pi )^2} \idx \rgh \ {15 \ov 2 \ep } \h R \nn \\ 
& + & {G \ov 4 \pi } \idx \rgh \ \bgl(
{15 \ov 4 \ep } \nab^2 \ol G_{\mn ,}{}^\mn | 
- {15 \ov 2 \ep } \nab_\m \nab_\n \ol G{^\n{}_{\r ,}{}^{\m \r }} | \nn \\
& &\qq \qq \qq + {6 \ov \ep } \h R_{\m \r \n \s } \ol G{^{\mn , \rs }} | 
+ {49 \ov 12 \ep } \h R_\mn \ol G{^\m{}_{\r ,}{}^{\n \r }} | 
- {31 \ov 24 \ep } \h R \ol G_{\mn ,}{}^\mn | \bgr) \ . 
\lbl{eqn:h1}
\eea

%The evaluation of the amplitudes of fig.\ref{fig:h}{\rm b} is much 
%easier than that of fig.\ref{fig:h}{\rm a}. It is straightforward from 
%the coincidence limits (\ref{eqn:limG}). We have two types of 
%expectation values to consider.
%\bea
%& & \idx \rgh \ <\nab_{\m } h_{\a \b } (x) \nab_{\n } h_{\g \dt } (x)> 
%<h_{\k \l } (x) h_{\ep \z } (x)>  \nn \\
%& = & 
%\idx \rgh \ 
%\nab_\m G_{\ab ,\g \dt } {\ola \nab}_{\n } | \ct 
%G_{\k \l ,\ep \z } | \ , \nn \\ 
%& & \idx \rgh \ <h_{\a \b } (x) h_{\g \dt } (x)> 
%<h_{\k \l } (x) h_{\ep \z } (x)> \h R_\mn (x) \nn \\
%& = & 
%\idx \rgh \ 
%G_{\ab ,\g \dt } | \ct 
%G_{\k \l ,\ep \z } | \ct \h R_\mn \ . 
%\eea

As for the amplitudes of fig.\ref{fig:h}{\rm b}, there are two types of the
expectation values to be considered. We have obtained the total divergent amplitude 
for fig.\ref{fig:h}{\rm b} as follows.
\bea
& & {G \ov (4 \pi )^2} \idx \rgh \ \bgl( - {11 \ov 6 \ep } \bgr) \h R \nn \\ 
& + & {G \ov 4 \pi } \idx \rgh \ \bgl(
- {5 \ov 6 \ep } \nab^2 \ol G_{\mn ,}{}^\mn | 
+ {5 \ov 3 \ep } \nab_\m \nab_\n \ol G{^\n{}_{\r ,}{}^{\m \r }} | \nn \\
&  &\qq \qq \qq \ -{1 \ov 3 \ep } \h R_{\m \r \n \s } 
\ol G{^{\mn , \rs }} | 
- {41 \ov 9 \ep } \h R_\mn \ol G{^\m{}_{\r ,}{}^{\n \r }} | 
+ {29 \ov 18 \ep } \h R \ol G_{\mn ,}{}^\mn | \bgr) \ . 
\lbl{eqn:h2}
\eea

%\vspace{1cm}

\subsection{Two-loop Diagrams with Ghost Propagators}

\begin{figure}
 \leavevmode
 \epsfxsize = 12 cm
 \centerline{ \epsfbox{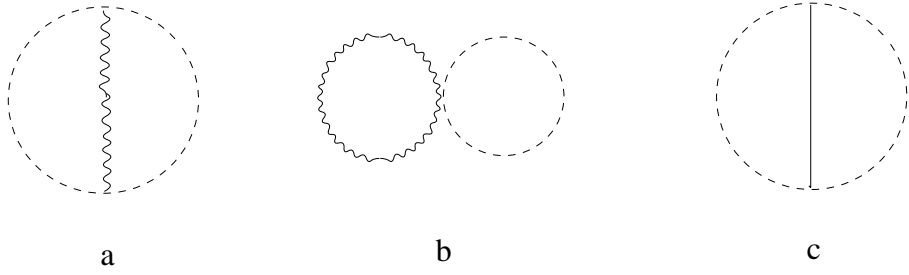}}
 \caption{Two-loop diagrams with ghost propagators}
 \label{fig:g}
\end{figure}

Thirdly, we consider the two-loop diagrams with the ghost propagators. 
In fig.\ref{fig:g}, the dashed lines denote the ghost propagators.

%In a similar way, we can evaluate the amplitudes of 
%fig.\ref{fig:g}{\rm a} and fig.\ref{fig:g}{\rm b} . 
We have three types of the diagrams figs.\ref{fig:g}{\rm a}-\ref{fig:g}{\rm c} 
for the ghost contribution. In a similar way, we can evaluate these amplitudes. 
The total divergent amplitude for fig.\ref{fig:g}{\rm a} is 
\bea
& & {G \ov (4 \pi )^2} \idx \rgh \ {11 \ov 4 \ep } \h R \nn \\ 
& + & {G \ov 4 \pi } \idx \rgh \ \bgl(
- {1 \ov 12 \ep } \nab^2 \ol G_{\mn ,}{}^\mn | 
+ {1 \ov 6 \ep } \nab_\m \nab_\n \ol G{^\n{}_{\r ,}{}^{\m \r }} | \nn \\
& &\qq \qq \qq \ 
+{1 \ov 3 \ep } \h R_{\m \r \n \s } \ol G{^{\mn , \rs }} | 
- {1 \ov 3 \ep } \h R_\mn \ol G{^\m{}_{\r ,}{}^{\n \r }} | 
- {1 \ov 12 \ep } \h R \ol G_{\mn ,}{}^\mn |  \nn \\
& &\qq \qq \qq \  -{1 \ov \ep } \nab^2 \ol G{^{\rm gh}{}_\m{}^\m} | 
+ {1 \ov \ep } \h R_\mn \ol G{^{{\rm gh} \ \m \n }} | 
\bgr) \ . 
\lbl{eqn:g1}
\eea

For fig.\ref{fig:g}{\rm b}, we have found the sum of the divergent amplitudes to be  
\bea
& & {G \ov (4 \pi )^2} \idx \rgh \ \bgl(- {14 \ov 9 \ep }\bgr) 
\h R \nn \\ 
& + & {G \ov 4 \pi } \idx \rgh \ \bgl(
- {1 \ov 3 \ep } \h R_{\m \r \n \s } \ol G{^{\mn , \rs }} | 
- {1 \ov 9 \ep } \h R_\mn \ol G{^\m{}_{\r ,}{}^{\n \r }} | 
+ {2 \ov 9 \ep } \h R \ol G_{\mn ,}{}^\mn |  \nn \\
& &\qq \qq \qq \  +{2 \ov 3 \ep} \nab^2 \ol G{^{\rm gh}{}_\m{}^\m} | 
- {2 \ov 3 \ep } \h R_\mn \ol G{^{{\rm gh} \ \m \n }} | 
\bgr) \ . 
\lbl{eqn:g2}
\eea

The amplitude for fig.\ref{fig:g}{\rm c} gives only the local single-pole 
divergence. This is also because the product of the three-point vertices 
is proportional to $a^2 \simeq {\ep / 2} $. The result is 
\be
{G \ov (4 \pi )^2} \idx \rgh \ \bgl(-{1 \ov 2 \ep } \bgr) \h R \ . 
\lbl{eqn:g3}
\ee
Again, we note that the result is obtained from the underlying 
double-pole singularity suppressed by $a^2 \simeq \ep /2 $. 

%For fig.\ref{fig:g}{\rm d}, the four point vertex is proportional 
%to $\ep$ and we have found no divergent contribution. 

%\vspace{1cm}

\subsection{Contribution from The $h_\mn$-$\ps$ Mixing Terms}

Here, we consider the diagrams with the mixings of $h_\mn$ and $\ps$ 
fields' propagators. 

\begin{figure}
 \leavevmode
 \epsfxsize = 10 cm
 \centerline{ \epsfbox{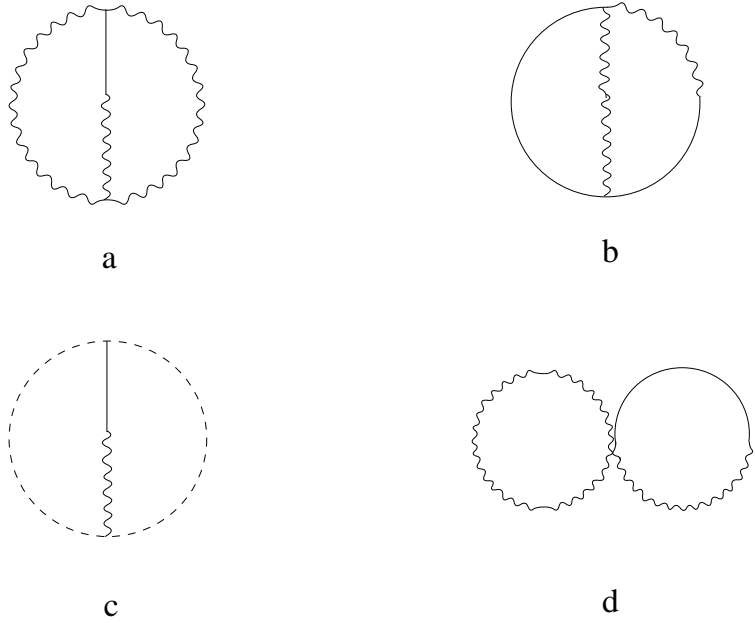}}
 \caption{Two-loop diagrams with the mixings of the $h_\mn$ and $\ps$ 
          fields' propagators}
 \label{fig:hpsi}
\end{figure}

There are four types of the diagrams 
figs.\ref{fig:hpsi}{\rm a}-\ref{fig:hpsi}{\rm d} 
which give nontrivial contributions. Each of them has only single mixing of 
$h_\mn$ and $\ps$ fields' propagators. The diagrams with the double 
mixings or more are found to give no divergent contribution. 

The evaluation of the diagrams with the mixings is much easier than that of 
the ones without the mixings, since there does not exist the parallel 
displacement matrix $a_0$ connecting $h_\mn$ field to $\ps$ 
field. We have obtained only the non-local divergences for each of 
the diagrams. This is because the local divergences necessarily arise from 
the coincidence limit (\ref{eqn:climhpsi}) of the $1$st Seeley 
coefficient $a_{1 \ \mn , \ps '}$ but it is a traceless matrix. 
As a result, we have found the sum of the amplitudes to be the following 
non-local divergences. 
\be
{G \ov 4 \pi } \idx \rgh \ {2 \sr{2} \ov 3 \ep} a i
(\nab_\m \nab_\n \ol G{^\mn{}_{, \ps}} | 
+ \h R_\mn \ol G{^\mn{}_{, \ps}} |) \ . 
\lbl{eqn:hpsi}
\ee
As it is expected they are just cancelled by the contribution from the 
counter terms (\ref{eqn:cterm}).

%\vspace{1cm}

\subsection{Contributions from The One-loop Finite or 
One-loop Counter Terms}

Finally, we evaluate the amplitudes for the diagrams with the insertion 
of the one-loop finite terms or the one-loop counter terms. 
\begin{figure}
 \leavevmode
\epsfxsize = 2.5 cm
 \centerline{ \epsfbox{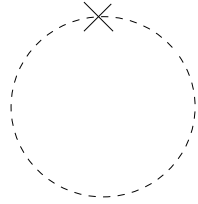}}
 \caption{Two-loop diagrams with the insertion of one-loop finite terms}
 \label{fig:f}
\end{figure}

We have the only diagram fig.\ref{fig:f}
to consider for the contribution from the one-loop finite terms. 
The divergent part of the diagram is found from 
(\ref{eqn:ga}) to be 
\be
{G \ov (4 \pi )^2} \idx \rgh \ {-25+c \ov 6 \ep } \h R \ .
\lbl{eqn:finite}
\ee

\begin{figure}
 \leavevmode
 \epsfxsize = 8 cm
 \centerline{ \epsfbox{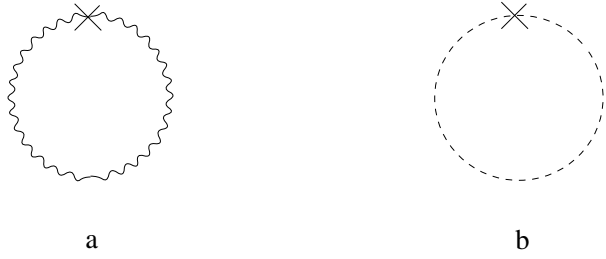}}
 \caption{Two-loop diagrams with the insertion of one-loop counter terms}
 \label{fig:c}
\end{figure}

In a similar fashion, we can easily evaluate the diagrams with the 
insertion of the one-loop counter terms. We have obtained the sum of 
the divergent amplitudes for figs.\ref{fig:c}{\rm a} and \ref{fig:c}{\rm b} 
from (\ref{eqn:limG}) and (\ref{eqn:cta}) as follows: 
\bea
& & {G \ov (4 \pi )^2} \idx \rgh \ {-95+3c \ov 9 \ep } \h R \nn \\ 
& + & {G \ov 4 \pi } \idx \rgh \ \bgl\{
- {33-c \ov 12 \ep } ( \nab^2 \ol G_{\mn ,}{}^\mn | 
- 2 \nab_\m \nab_\n \ol G{^\n{}_{\r ,}{}^{\m \r }} | 
+ 2 \h R_{\m \r \n \s } \ol G{^{\mn , \rs }} | ) \nn \\ 
& &\qq \qq \qq \ - {2 \sr{2} \ov 3 \ep} a i
(\nab_\m \nab_\n \ol G{^\mn{}_{, \ps}} | 
+ \h R_\mn \ol G{^\mn{}_{, \ps}} |) \nn \\
& &\qq \qq \qq \  
+ {1 \ov 3 \ep } (\nab^2 \ol G{^{\rm gh}{}_\m{}^\m} | 
- \h R_\mn \ol G{^{{\rm gh} \ \m \n }} | ) 
\bgr\} \ . 
\lbl{eqn:cterm}
\eea

%\vspace{1.5cm}

\subsection{Total Amplitude}

After these calculations, the sum of the two loop divergences except 
those from the counter terms is evaluated by summing the
contributions of 
(\ref{eqn:psi1}), (\ref{eqn:psi2}), 
(\ref{eqn:psi3}), (\ref{eqn:h1}), (\ref{eqn:h2}), (\ref{eqn:g1}), 
(\ref{eqn:g2}), (\ref{eqn:g3}), (\ref{eqn:hpsi}) and (\ref{eqn:finite}) :
\bea
& & {G \ov (4 \pi )^2} \idx \rgh \ {113+3c \ov 72 \ep } \h R \nn \\ 
& + & {G \ov 4 \pi } \idx \rgh \ \bgl\{ 
{33-c \ov 12 \ep } ( \nab^2 \ol G_{\mn ,}{}^\mn | 
- 2 \nab_\m \nab_\n \ol G{^\n{}_{\r ,}{}^{\m \r }} | 
+ 2 \h R_{\m \r \n \s } \ol G{^{\mn , \rs }} | ) \nn \\ 
& &\qq \qq \qq + {2 \sr{2} \ov 3 \ep} a i
(\nab_\m \nab_\n \ol G{^\mn{}_{, \ps}} | 
+ \h R_\mn \ol G{^\mn{}_{, \ps}} |) \nn \\
& &\qq \qq \qq - {1 \ov 3 \ep } (\nab^2 \ol G{^{\rm gh}{}_\m{}^\m} | 
- \h R_\mn \ol G{^{{\rm gh} \ \m \n }} | ) \nn \\ 
& &\qq \qq \qq - {11 \ov 12 \ep } (\h R_\mn - \ah \h g_\mn \h R) \  
\ol G{^\m{}_{\r ,}{}^{\n \r }} | \  
\bgr\} \ . 
\lbl{eqn:tot1}
\eea
Here, we note that the non-local divergences originate from the one-loop 
sub-divergences multiplied by the non-local parts of the propagators 
with possible derivatives $\ol G, \nab^2 \ol G, \nab_\m \nab_\n \ol G$. 
This is directly seen from our calculation procedure. When we evaluate 
a non-local divergence at the two-loop level, we replace one of the 
propagators of a two-loop diagram by its non-local part $\ol G$. 
A non-local divergence is just a non-local part $\ol G$ multiplied 
by a local divergence of a sub-loop which exist in the relevant
diagram. 

Therefore the non-local divergences at the two loop level exactly match 
the one-loop divergent corrections of quantum two-point 
functions. For example, let us consider a pair of the three-point vertices 
of quantum fields. We choose a pair of quantum fields from the 
each of the vertices 
as external fields. Next, we make a loop connecting  
the remaining two-point vertices. The one-loop divergent correction 
of a quantum 
two-point function is the two external fields multiplied by the local 
divergence of the loop. This is identical to a non-local divergence 
of a two-loop diagram, if we replace the two external fields by 
the non-local part of their propagator $\ol G$. 
Conversely we have the following transformation rules, which give a one-loop 
divergence for a quantum two-point function from a non-local divergence 
of a two-loop diagram. For example, 
\bea
{{\rm const.} \ov \ep} \nab_\m \nab_\n \ol G{^\n{}_{\r ,}{}^{\m \r }} | 
& \longrightarrow & - {{\rm const.} \ov \ep} 
\nab_\m h^\m{}_\r \nab_\n h^{\n \r} \ , \nn \\ 
{{\rm const.} \ov \ep} \h R_{\m \r \n \s } \ol G{^{\mn , \rs }} | 
& \longrightarrow & {{\rm const.} \ov \ep} 
\h R_{\m \r \n \s } h^\mn h^\rs \ . 
\eea

We can therefore determine the one-loop quantum field renormalizations 
which are required to cancel the nonlocal divergences of the
two loop amplitudes
if we evaluate the non-local parts of the two-loop corrections. 
The renormalizability of the theory implies that the necessary
counter terms can be supplied by the renormalization of the
couplings and the fields.
On the other hand, we have already determined the counter terms (\ref{eqn:cta}) 
which consist of 
the one-loop renormalization of the gravitational coupling constant 
and the linear wave function renormalization of the quantum fields\cite{AKKN}. 

The sum of the amplitudes (\ref{eqn:tot1}) 
and (\ref{eqn:cterm}) are indeed found to be local except
the term which is proportional to the field equation: 
\bea
& & {G \ov (4 \pi )^2} \idx \rgh \ 
{-79 + 3 c \ov 8 \ep } \h R \nn \\ 
& + & {G \ov 4 \pi } \idx \rgh \ 
\bgl(- {11 \ov 12 \ep } \bgr) (\h R_\mn - \ah \h g_\mn \h R) 
\ol G{^\m{}_{\r ,}{}^{\n \r }} | \ . 
\lbl{eqn:tot2}
\eea
Since the field equations appear when we take the variation
of the action by the fields, the remaining divergence can be dealt with
by a nonlinear field renormalization at the one loop level.
Previously we missed this counter term since we considered only
the linear wave function renormalization of the fields.
We remark that there is no principle to prohibit the nonlinear
field renormalizations in this theory since
the fields are dimensionless.
Therefore, we are led to introduce the following renormalization of 
$h_\mn$ field, paying attention to the traceless property. 
\be
h^0_\mn = \bgl( 1 - {G \ov 6 \pi \ep} \bgr) h_\mn \ + \ 
Z_{\rm n.l.} \bgl( h_{\m \r} h^\r{}_\n 
- {1 \ov D} \h g_\mn h_\rs h^\rs \bgr) \ . 
\lbl{eqn:nonlin}
\ee
From the linear terms of the expansion (\ref{eqn:expR}) we obtain 
additional counter terms, and find the expectation values of them 
to give only the non-local divergences: 
\be
\idx \rgh \ 2 Z_{\rm n.l.} (\h R_\mn - \ah \h g_\mn \h R) 
\ol G{^\m{}_{\r ,}{}^{\n \r }} | \ . 
\ee
Therefore, the two-loop expectation value (\ref{eqn:tot2}) can be made 
to be local, only if we choose the non-linear renormalization 
constant $Z_{\rm n.l.}$ of $h_\mn$ field as
\be
Z_{\rm n.l.} \ = \ {G \ov 4 \pi} {11 \ov 24 \ep} \ . 
\ee

%Finally, to remove the non-local divergences in (\ref{eqn:tot1}), 
%we use the relations obtained from (\ref{eqn:nl}).
%\bea
%(\h g_{\m \r } \h g_{\n \s } \nab^2 + 2 \h R_{\m \r \n \s }) 
%\ol G^{\mn , \rs } | & \simeq & 
%{1 \ov 4 \pi } \bgl(- {5 \ov 3} \bgr) \h R \ , \nn \\
%(\h g_\mn \nab^2 - \h R_\mn ) 
%\ol G^{{\rm gh} \ \mn } | & \simeq & 
%{1 \ov 4 \pi } \bgl({4 \ov 3} \bgr) \h R \ .
%\eea

Finally the total divergent contribution to the effective action
at the two loop level in pure Einstein gravity is found to be of the 
renormalizable form. 
\be
{G \ov (4 \pi )^2} \idx \rgh \ 
{-79 + 3 c \ov 8 \ep } \h R \ . 
\lbl{eqn:tot3}
\ee
Here, we note that there arises only the single-pole divergence. 
The cancellation of the double-pole divergences is necessary for
the consistency of the theory.

\section{Renormalization of the Cosmological Constant Operator}
\setcounter{equation}{0}
\lbl{sec:cosmo}

In this section, we study the renormalization of the cosmological
constant operator.
As it will be shown, we need to introduce new counter terms 
to renormalize the cosmological constant operator.
We also show that these new counter terms do not spoil 
the result of the previous section.
We consider an infinitesimal perturbation of the theory by 
the cosmological constant operator by adding the 
following term to the action:
\bea
\Lambda \idx \sr{g} & = & \Lambda \idx \sr{\h{g}}(1 + {2 \ep b \ov a} \ps )^{2D \ov \ep} 
\nn \\ 
& \simeq & \Lambda \idx \sr{\h{g}}\exp \bl( (1-{\ep\ov 2}){1 \ov a} \ps - {\ep \ov 8 a^2}
 \ps^2 
+ \lts \br) \ . 
\lbl{eqn:cos}
\eea
This term is manifestly generally covariant and it is invariant under 
the gauge transformation (\ref{eqn:gtr1}).
The quadratic part of the action with respect to $\psi$ field becomes:
\be
\idx \sr{\h{g}}\bgl( -{\mu^{\ep}\ov 2G}(1-a^2)\partial _{\mu}\psi\partial ^{\mu}\psi
+\Lambda \mu^D (1-{5\ep\ov 4}){1 \ov 2 a^2}\psi ^2 +{1\ov G}\ep b \hat{R} \psi ^2\bgr) \ .
\ee

We regard $\Lambda$ to be infinitesimally small and evaluate
the divergent part of the effective action proportional to $\Lambda$.
Namely we consider the diagrams with the single insertion of the
cosmological constant operator.
Further insertions of the cosmological constant operator
do not lead to the short distance divergences by the power counting.
We utilize the same covariant calculation method which has been 
employed in the previous section.
The relevant diagrams in this section are those which contain $\psi$ field
propagators.
The divergent part proportional to $\Lambda$ at the one loop level is
\be
-\idx \sr{\h{g}}({G\over {4\pi }})\Lambda ({1\ov \ep a^2} +{1\ov \ep}) \ .
\lbl{eqn:cos1}
\ee
At the two loop level, we find the following divergences:
\be
\idx \sr{\h{g}}
\bgl( ({G\over {4\pi }})^2\Lambda 
\bl( ({1\over {\epsilon a^2}})({-1\over 4})+
{-3\ov \ep ^2}+({1\ov \ep })({-35\ov 8}+4\pi \bar{G}_{\mu\nu},^{\mu\nu})
\br)
+{G\over {4\pi \ep }} a^2  \nab ^2\bar{G}_{\psi\psi}|\bgr) \ .
\lbl{eqn:cos2}
\ee
Here we omit the square of the one loop result (\ref{eqn:cos1}).
In this expression, we find two types of the nonlocal divergences.
%In this expression, we find a nonlocal divergence.

The first type arises due to the one loop level divergence of  
${G\over {4\pi\ep }}\Lambda \idx \sr{\h{g}} h_{\mu\nu}h^{\mu\nu}$ form.
It also leads to the double pole singularity in (\ref{eqn:cos2}).
Such a divergence can be subtracted by the nonlinear wave function 
renormalization of $\psi$ field as
$\psi ^0 = \psi +a ({G\over {4\pi \ep}}) (h_{\mu\nu}h^{\mu\nu}+{G\over \pi\ep})$.
The last term at $O(G^2)$ is added in order to cancel the leading singularity
of the vacuum expectation value of $\psi$.
By doing so, we can eliminate the first nonlocal divergence in (\ref{eqn:cos2})
and obtain
\be
\idx \sr{\h{g}}\bgl( ({G\over {4\pi }})^2\Lambda 
\bl( ({1\over {\epsilon a^2}})({-1\over 4} )+
{-3\ov \ep ^2}+({1\ov \ep })({-27\ov 8} )\br) 
+{G\over {4\pi \ep }} a^2  \nab ^2\bar{G}_{\psi\psi}|\bgr)\ .
\ee
This procedure also gives rise to the following counter term
from the Einstein action:
\be
({a^2\ov 4\pi\ep})
\idx \sr{\h{g}}\h{R}(h_{\mu\nu}h^{\mu\nu}+{G\over \pi\ep}) \ .
\ee
At the one loop level, it can be regarded to be a finite counter
term by expanding $a^2$ in terms of $G$. It leads to no new divergence
of $\idx \sr{\h{g}}\h{R}$ type at the two loop level.
Therefore this new counter term is harmless to the order
in the perturbation theory we are working here.

We have also found the following contribution
of the $\psi$ field kinetic term type due to the one loop
subloops of $h_{\mu\nu}$ and ghost fields
which is responsible for the
second nonlocal divergence in (\ref{eqn:cos2}):
\be
-({a^2\ov 4\pi \ep}) \idx \sr{\h{g}}
\partial _{\mu} \psi\partial ^{\mu} \psi \ .
\ee
%It also leads to a nonlocal divergence in (\ref{eqn:cos2})
%although we have suppressed it. 
This one loop contribution is another source of the double
pole singularity of the cosmological constant operator at
the two loop level.
It is because $a^2$ factor in the numerator of this term can
be cancelled by the inverse power of $a$ which is associated
with $\psi$ field in the cosmological constant operator.
We need to subtract this term from the one loop subdiagrams
when we consider the diagrams with the single insertion of
the cosmological constant operator.
%Although it is finite to the leading order, we choose
%to deal with it by the following procedure.
In order to do so,
we have chosen to adopt
the following counter term in the action:
\be
({(D-1)a^2\ov 2\pi \ep})
\idx \sr{\h{g}} \bgl( \tilde{R}(1+a\psi+\ep b\psi ^2)-
{1\ov 2}\partial _{\mu} \psi\partial ^{\mu} \psi\bgr) \ .
\lbl{eqn:countf}
\ee
This action contains the desired counter term of the $\psi$ field 
kinetic term type. Although it 
contains other counter terms, they can be regarded as
finite counter terms by expanding $a^2$ in terms of $G$.
It is because $\psi$ field appears with additional 
factors of $a$ or $\epsilon$ as long as they are concerned. 
The terms which involve
only $h_{\mu\nu}$ field are safe since they do not
couple to $\psi$ field directory.
After this procedure, the two loop level divergence
proportinal to $\Lambda$ becomes:
\be
\idx \sr{\h{g}}({G\over {4\pi }})^2\Lambda 
\bgl( ({1\over {\epsilon a^2}})({-1\over 4} )+
{-1\ov \ep ^2}+({1\ov \ep })({-7\ov 8} )\bgr) \ .
\ee
The counter terms (\ref{eqn:countf}) do not give rise to
new divergences of the Einstein action form neither at the two loop level.

After these considerations, we find that the divergence of the effective
action which is proportinal to $\Lambda$ is local.
However it still contains the double pole in $\ep$.
The double pole here can be removed by the following wave function
renormalization. 
We note that there is a freedom of the scale transformation of the background
metric. 
We utilize this freedom to eliminate the
divergences of the cosmological constant type which
is independent of $a^2$.  
We rescale the background metric by the following factor:
\be
Z=1
+({G\over {4\pi }})^2 
({-1\ov \ep ^2}) \ .
\ee
After the scaling, the remaining
two loop level divergence proportinal to $\Lambda$ has only the 
simple pole in $\ep$:
\be
\idx \sr{\h{g}}({G\over {4\pi }})^2\Lambda 
\bgl( ({1\over {\epsilon a^2}})({-1\over 4})+({1\ov \ep })({-7\ov 8} )\bgr) \ .
\ee

This procedure also leads to the rescaling of 
the Einstein action. 
\bea
& & Z^{{\ep \ov 2}} \idx \rgh {1 \ov G} \hat{R} \ , \nn \\ 
& & Z^{{\ep \ov 2}} \ = \ 
1 
+\bgl( {G \ov 4 \pi} \bgr)^2 ({-1 \ov 2 \ep})\ .
\lbl{eqn:rescale}
\eea
Here, the exponent: ${\ep \ov 2}$ is seen from the dependence of the 
Einstein term and the cosmological constant operator on the conformal 
mode. 
\be
\sr{g} R \simeq e^{- {\ep \ov 2} \p}  \rgh\til R \ \ , \ \ 
\sr{g} \simeq e^{- {D \ov 2} \p} \rgh \ . 
\ee
The rescaling of the background metric introduces a new
counter term as in (\ref{eqn:rescale}). 
Due to the new counter term, the total two-loop divergence 
of the effective action which is of the Einstein action type becomes
the following instead of the final result of the previous section (\ref{eqn:tot3}):
\be
{G \ov (4 \pi )^2} \idx \rgh \ 
{ -3(25- c) \ov 8 \ep } \h R \ . 
\lbl{eqn:tot4}
\ee
It has a simple pole in $\ep$ and vanishes when $c=25$ which
corresponds to the critical string theory in the two
dimensional limit. These are certainly the desirable features.

Here we recall the counter terms at the one loop level
which can be associated with the bare
gravitational coupling constant
\bea
& & -{\mu ^\ep\ov G}({25-c\ov 6\ep})({G\ov 4\pi})
+{(D-1)a^2\ov 2\pi\ep},\nn \\
& = & -{\mu ^\ep\ov G}({25-c\ov 6\ep})({G\ov 4\pi})
+{\mu^\ep\ov 4\pi}\bgl( 1-({25-c\ov 6\ep})({G\ov 4\pi})\bgr)
\eea
The first term is the coefficient of $\idx \sr{\h{g}} \tilde{R}$ and the
second term is the coefficient of the Einstein action in
(\ref{eqn:countf}).
The only first term contributes to the conformal anomaly
at the one loop level and enters in the definition
of $a^2$ at $O(G)$\cite{KKN3}.
We note that the second term contains $O(G)$ term.
We have to add an appropriate $O(G)$ term to it in order
to cancel (\ref{eqn:tot4}).

In this way,
the bare gravitational coupling constant 
at the two loop level is found to be
\be
{1\ov G^0} = {\mu ^\ep\ov G}(1+{G\ov 4\pi})
\bgl( 1-({25-c\ov 6\ep})({G\ov 4\pi})-
({5(25-c)\ov 24\ep})({G\ov 4\pi})^2\bgr)
\ee
We can still remove the factor $(1+{G\ov 4\pi})$ from this quantity 
by rescaling the background
metric. We also need to multiply the factor $(1+{G\ov 4\pi})^{-2\ov \ep}$ to the
cosmological constant operator at the same time.
In such a renormalization scheme, the bare gravitational coupling constant
is found to be:
\be
{1\ov G^0} = {\mu ^\ep\ov G}
\bgl( 1-({25-c\ov 6\ep})({G\ov 4\pi})-
({5(25-c)\ov 24\ep})({G\ov 4\pi})^2\bgr)
\ee
The $\beta$ function of the gravitational coupling constant is
\be
\beta _G = \ep G-({25-c\ov 6})({G\ov 4\pi})G-
({5(25-c)\ov 12})({G\ov 4\pi})^2G
\ee
It agrees with our previous result of \cite{AKNT} to the leading order in $c$.

The remaining simple pole divergences
of the cosmological constant operator which are
inversely proportional to $a^2$ can be ascribed
to the singular expectation value of $\psi ^2$: 
\be
<\psi ^2> = {G\ov 2\pi\ep}\bgl( 1+({1\ov 4}){G\ov 4\pi}\bgr)
\ee
They are of order $(G/a^2 )$ and 
they can become large around the short distance fixed point
of the renormalization group where $a^2 \rightarrow 0$.
At higher orders, we also find the divergences of order $(G/a^2 )^n$.
Therefore we need to sum this class of contributions to all
orders. 
It can be done by considering the following integral\cite{KKN1}.
\be
\int d \psi exp\bgl( {4\ov \ep}log(1+{\ep\psi\ov 4a})-{\pi \ep\ov G}(
1-({1\ov 4}){G\ov 4\pi})\psi ^2\bgr)
\ee
This integral can be evaluated for small $\ep$ by the saddle point method
with the change of the variables: ${\ep\psi\ov 4a}=\rho$.
The saddle point of $\rho$ satisfies:
\be
\rho _0(1+\rho _0)-{G\ov 8\pi a^2}\bgl( 1+({1\ov 4}){G\ov 4\pi}\bgr)
=0
\ee
The renormalization factor is found to be
\be
Z_{\Lambda} = exp\bgl( {4\ov \ep}log(1+\rho _0)-{16\pi a^2\ov \ep G}(
1-{G\ov 16\pi})\rho _0 ^2\bgr) exp(-{G\ov 4\pi\ep})
\ee
where the last factor is necessary to cancel the one loop
divergence which is independent of $a^2$.
The anomalous dimension is found to be
\bea
\gamma &=& \mu {\partial \ov \partial \mu}log Z_{\Lambda} \nn,\\
&=& {16\pi a^2\ov G} \rho _0^2 -{G\ov 4\pi}
\eea
We therefore find the anomalous dimension to be
\be
2(1-{G\ov 16\pi})-{8\pi\ov G}
\bgl( -a^2+\sr{a^4+{Ga^2\ov 2\pi}(1+{G\ov 16\pi})}\bgr)
\ee
In these considerations, $a^2$ is regarded as $O(G)$ quantity and 
the anomalous dimension is $O(1)$ at the one loop level.
At the two loop level, we find a finite correction of $O(G)$
to the leading result.

At the short distance fixed point of the renormalization group where 
$a^2$ vanishes,
we find the anomalous dimension of the cosmological constant operator
to be
\be
2(1-{G^*\ov 16\pi})
\ee
where $G^*={24\pi \ep\ov 25-c} +O(\ep^2 )$ is the fixed point coupling constant.
The scaling dimension is the sum of the canonical and the anomalous dimensions.
The scaling dimension of the cosmological constant operator around the fixed point
is found to be
\be
-(2+\ep) + 2-{G^*\ov 8\pi}
= -\ep-{3\ep\ov 25-c}+O(\ep^2 )
\ee
The leading order quantity of $O(1)$ has been cancelled by the
quantum correction and the scaling dimension is found to be 
$O(\ep)$ by the two loop level calculation.

The scaling dimension of the inverse gravitational coupling constant
around the fixed point is:
\be
\mu{\d\ov \d\mu}({1\ov G}-{1\ov G^*}) =-\ep({1\ov G}-{1\ov G^*})+O(\ep^2 )
\ee
Since the scaling dimension of the cosmological constant operator
is determined to be $O(\ep)$ by the two loop calculation,
we only need the one loop $\beta$ function of the gravitational coupling
constant here.
It is because of the cancelation of the leading $O(1)$ term in the
scaling dimension of the cosmological constant operator.
In order to determine the scaling dimension up to $O(\ep ^2)$,
we need to renormalize the cosmological constant operator
up to higher orders.
We find that the both gravitational and the cosmological couplings are relevant. 
Although they are also
relevant couplings classically, the scaling dimension of the
cosmological constant operators has changed drastically from the classical
value of $D$ to that of $O(\ep )$.
However these scaling dimensions separately cannot be physical
quantities. It is because they are scheme dependent on how
we rescale the metric.

The physically meaningful prediction is the scaling relation
between the gravitational
coupling constant and the cosmological constant.
We find the scaling relation between the gravitational
coupling constant and the cosmological constant 
at the short distance fixed point as
\be
({1\ov G}-{1\ov G^*})^{\ep+{3\ep\ov 25-c}} \sim \Lambda^{\ep}
\lbl{eqn:scale}
\ee
It is very different from  
the classical scaling relation which holds at the weak coupling 
infrared fixed point:
\be
({1\ov G})^{D} \sim \Lambda^{\ep}
\ee

These scaling exponents have been numerically measured 
in four dimensions by using a real space renormalization group
technique\cite{BKK}.
What has been observed is the following scaling relation
at an ultraviolet stable fixed point:
\be
N^{-{\beta _0\ov 4}} \sim |\kappa ^* -\kappa |
\ee
where $\kappa$ is the inverse gravitational coupling
constant and $N$ is the space time volume which is dual to
the cosmological constant. Therefore it leads
\be
\Lambda^{{\beta _0\ov 4}} \sim |\kappa ^* -\kappa |
\ee
$\beta _0$ is measured to be $6.5(1.2)$.
If we put $\ep =2$ in our scaling relation (\ref{eqn:scale}), 
it gives $\beta _0 \sim 4$.
On the other hand, the classical scaling relation corresponds to the
exponent $\beta _0 = 2$.
We observe that our prediction is closer to the measured value than 
the classical exponent.

\section{Conclusions and Discussions}
\lbl{sec:con}

In this paper, we have performed the two loop level renormalization
of the quantum gravity in $2+\ep$ dimensions. 
We have adopted the background gauge and computed the effective
action of the background fields which is manifestly covariant.
We have made the most of the covariance to simplify
the calculation.
We have shown that
the short distance divergences are renormalizable within our scheme
to the two loop level.

In fact we have explicitly determined the counter terms which
make the effective action finite up to this order.
These counter terms are all of the expected types which can be
supplied by the renormalization of the couplings and the
wave functions. The most nontrivial counter term is the type of
$\idx \sr{\h{g}} \tilde{R}$ which gives rise to the conformal
anomaly. Our renormalization scheme is successful
to handle it by choosing the ``linear dilaton" type coupling $a$ 
in such a way to maintain
the conformal invariance with respect to the background metric.
The bare action can be chosen to be a manifestly covariant
form simultaneously\cite{KKN3}.
We  have thus demonstrated the effectiveness of our formalism
by the explicit calculations reported here.

We have further predicted the scaling relations and the scaling exponents 
between the gravitational and the cosmological coupling constants.
Such a prediction can be compared with recent numerical
investigations in quantum gravity. 
We have found that our prediction appears to be closer
to the measured value than the classical prediction.
However it also turns out that taking the continuum
limit is not straightforward at the ultraviolet stable fixed
point where the scaling is observed\cite{BBKP}.
The problem is that the phase transition appears to be of the first order
rather than the second order.
So constructing a realistic quantum gravity theory
still is a challenging task and the
exponent which has been measured in the numerical approach
may be taken qualitatively at best. 

However we cannot rule out the possibility that the continuum
limit may be taken with further ingenuity
and the scaling behavior currently observed 
%in the simplical quantum gravity may 
may qualitatively survive.
Therefore we believe that it is  worthwhile to seriously attempt to
calculate the scaling exponents in the
continuum approach
in view of the progress
which has been already achieved in the numerical approach.
We hope that our results will shed light on the continuum quantum
gravity with such an optimism.

%\vspace{1cm}

\section*{Acknowledgements}
\addcontentsline{toc}{section}{\protect\numberline{}{Acknowledgements}}
We thank H. Kawai ,M. Ninomiya and J. Ambjorn for stimulating discussions.
We also thank J. Nishimura and A. Tsuchiya for the earlier collaboration
on this project.
We have performed a large amount of tensor calculations with the help 
of MathTensor, a software for symbolic manipulations involved in tensor 
analysis. 
Our work is supported in part by the Grant-in-Aid for Scientific Research
from the Ministry of Education,Science and Culture of Japan.

%\newpage
%\vspace{1cm}

\appendix
%\section{Conventions}
%\lbl{sec:cc}

%In this paper, we have adopted the curvature conventions 
%as follows.
%\be
%[\nab_\m , \nab_\n ] A^\r \ = \ R^\r{}_{\s \m \n} A^\s , \ \
%R_\mn \ = \ R^\r{}_{\m \n \r} , \ \ 
%R \ = \ g^\mn R_\mn .
%\ee

%\vspace{1cm}

\section{Singular Products of The Functions $G_0 , G_1$} 
\setcounter{equation}{0}
\lbl{sec:sin}

In the course of the calculation of the two-loop amplitudes, 
we need to know the singular behavior of the several products of the 
functions $G_0$'s and $G_1$'s at short distance. In general, 
the singular part is given 
by the bi-scalar delta functions with derivatives or those multiplied 
by covariant curvature tensors. In terms of them, we can evaluate the loop 
amplitudes in a manifestly covariant way. In this appendix, we will 
illustrate the derivation of them \cite{O} . 

The functions $G_0 $ and $G_1 $ are effectively defined by the 
equations (\ref{eqn:rel1}). The consistency with the first 
of (\ref{eqn:rel1}) and the coincidence limits (\ref{eqn:limG0}) 
lead us to find the singular behavior:
\be
\nab_\m \nab_\n G_0 \ct G_0{}^n \sim 
- {(-1)^n \ov D} {\m^{n \ep } \ov (2 \pi \ep )^n } \ \h g_\mn \dt^D \ , 
\lbl{eqn:sin1}
\ee
from which we can immediately obtain, discarding a non-singular total 
derivative, 
\be 
\nab_\m G_0 \nab_\n G_0 \ct G_0{}^{n-1} \sim 
{1 \ov D} {(-1)^n \ov n} {\m^{n \ep } \ov (2 \pi \ep )^n } 
\ \h g_\mn \dt^D \ .
\lbl{eqn:sin2}
\ee

%\vspace{0.5cm}

Differentiating (\ref{eqn:sin1}) and (\ref{eqn:sin2}), we can derive the 
singular behavior of the products with three derivatives. They are 
proportional to the delta functions with a derivative, since there does not 
exist a covariant curvature term of the mass dimension one. 
\bea
& & \nab_\m \nab_\n G_0 \nab_\r G_0 \ct G_0{}^{n-1} \nn \\
& \sim & - {1 \ov D+2} {(-1)^n \ov n} {\m^{n \ep } \ov (2 \pi \ep )^n } 
\bgl( {2 \ov D} \h g_\mn \d_\r - \h g_{\m \r } \d_\n 
- \h g_{\n \r } \d_\m \bgr) \dt^D \ , 
\lbl{eqn:sin3} \\ 
& & \qq \nn \\ 
& & \nab_\m G_0 \nab_\n G_0 \nab_\r G_0 \ct G_0{}^{n-2} \nn \\
& \sim & - {\ep \ov D (D+2)} {(-1)^n \ov n (n-1)} 
{\m^{n \ep } \ov (2 \pi \ep )^n } 
( \h g_\mn \d_\r + \h g_{\m \r } \d_\n + \h g_{\n \r } \d_\m ) \dt^D \ . 
\lbl{eqn:sin4}
\eea
Here, we have used the result.
\bea
& & \nab_\m \nab_\n \nab_\r G_0 \ct G_0{}^n \nn \\
& \sim & - {(-1)^n \ov D+2} {\m^{n \ep } \ov (2 \pi \ep )^n } 
( \h g_\mn \d_\r + \h g_{\m \r } \d_\n + \h g_{\n \r } \d_\m ) \dt^D \ , 
\lbl{eqn:sin5}
\eea
which is obtained, in the same way as (\ref{eqn:sin1}), from the symmetry 
under the exchange of the indices of the singular part. 

%\vspace{0.5cm}

Next, we proceed to the singular behavior of the products of 
the functions $G_0$'s and $G_1$'s with four derivatives. They have 
the mass dimension four. So, the singular piece consists of the 
bi-scalar delta functions with the quadratic derivatives or those 
multiplied by the curvature tensors. 

We cite the results for the singular parts of the products 
of $G_0$'s and $G_1$'s. 
\bea
& & \nab_\m \nab_\n G_0 \nab_\r \nab_\s G_0 \ct G_0{}^{n-1} \nn \\
& \sim & 
{1 \ov (D+2) (D+4)} {(-1)^n \ov n} {\m^{n \ep } \ov (2 \pi \ep )^n } 
\bgl\{ {2- \ep \ov D} \h g_\mn \h g_\rs \nab^2 \dt^D 
+ {D \ov 2} (\h g_{\m \r} \h g_{\n \s} + \h g_{\m \s} \h g_{\n \r}) 
\nab^2 \dt^D \nn \\ 
& - & 4 (\h g_\mn \nab_\r \nab_\s + \h g_\rs \nab_\m \nab_\n ) \dt^D 
\nn \\ 
& + & D (\h g_{\m \r } \nab_\n \nab_\s 
+ \h g_{\m \s } \nab_\n \nab_\r + \h g_{\n \r } \nab_\m \nab_\s 
+ \h g_{\n \s } \nab_\m \nab_\r ) \dt^D \bgr\} \nn \\  
& + & {1 \ov 3 (D+2)} {(-1)^n \ov n} {\m^{n \ep } \ov (2 \pi \ep )^n } 
\{ 2 ( \h R_{\m \r \n \s } + \h R_{\m \s \n \r } ) \nn \\ 
& + & (\h g_{\m \r } \h R_{\n \s } + \h g_{\m \s } \h R_{\n \r } + 
\h g_{\n \r } \h R_{\m \s } + \h g_{\n \s } \h R_{\m \r } ) \} 
\dt^D \nn \\
& - & {1 \ov 6 (D+2) (D+4)} {(-1)^n (n-1) \ov n} 
{\m^{n \ep } \ov (2 \pi \ep )^n } 
\bgl\{ \bgl(1 - {2 \ep \ov D} \bgr) \h g_\mn \h g_\rs \h R \nn \\ 
& + & {D \ov 2} (\h g_{\m \r } \h g_{\n \s } + \h g_{\m \s } \h g_{\n \r }) 
\h R - 4 (\h g_\mn \h R_\rs + \h g_\rs \h R_\mn ) \nn \\ 
& + & D (\h g_{\m \r } \h R_{\n \s } + \h g_{\m \s } \h R_{\n \r } + 
\h g_{\n \r } \h R_{\m \s } + \h g_{\n \s } \h R_{\m \r } ) \bgr\} 
\dt^D \nn \\
& + & {(-1)^n \ov 6 D} {\m^{n \ep } \ov (2 \pi \ep )^n } 
(\h g_\mn \h R_\rs + \h g_\rs \h R_\mn ) \dt^D \ , 
\lbl{eqn:sin6} \\ 
& & \qq \nn \\ 
& & \nab_\m \nab_\n G_0 \nab_\r G_0 \nab_\s G_0 
\ct G_0{}^{n-2} \nn \\
& \sim & 
{\ep \ov (D+2) (D+4)} {(-1)^n \ov n (n-1)} {\m^{n \ep } \ov (2 \pi \ep )^n } 
\bgl\{ {2 \ov D} \h g_\mn \h g_\rs \nab^2 \dt^D 
- \ah (\h g_{\m \r} \h g_{\n \s} + \h g_{\m \s} \h g_{\n \r}) 
\nab^2 \dt^D \nn \\ 
& + & {4 \ov D} \h g_\mn \nab_\r \nab_\s \dt^D 
- \h g_\rs \nab_\m \nab_\n \dt^D \nn \\  
& - & (\h g_{\m \r } \nab_\n \nab_\s 
+ \h g_{\m \s } \nab_\n \nab_\r + \h g_{\n \r } \nab_\m \nab_\s 
+ \h g_{\n \s } \nab_\m \nab_\r ) \dt^D \bgr\} \nn \\
& - & {\ep \ov 3 D (D+2)} {(-1)^n \ov n (n-1)} 
{\m^{n \ep } \ov (2 \pi \ep )^n } 
\{ 2 ( \h R_{\m \r \n \s } + \h R_{\m \s \n \r } ) \nn \\ 
& + & (\h g_{\m \r } \h R_{\n \s } + \h g_{\m \s } \h R_{\n \r } + 
\h g_{\n \r } \h R_{\m \s } + \h g_{\n \s } \h R_{\m \r } ) \} 
\dt^D \nn \\
& + & {\ep \ov 6 (D+2) (D+4)} {(-1)^n \ov n} 
{\m^{n \ep } \ov (2 \pi \ep )^n } 
\bgl\{ - {2 \ov D} \h g_\mn \h g_\rs \h R \nn \\ 
& + & \ah (\h g_{\m \r } \h g_{\n \s } + \h g_{\m \s } \h g_{\n \r }) 
\h R + (\h g_\mn \h R_\rs + \h g_\rs \h R_\mn ) \nn \\ 
& - & {D \ov 4} (\h g_{\m \r } \h R_{\n \s } + \h g_{\m \s } \h R_{\n \r } 
+ \h g_{\n \r } \h R_{\m \s } + \h g_{\n \s } \h R_{\m \r } ) \bgr\} 
\dt^D \nn \\
& - & {(-1)^n \ov 6 n} {\m^{n \ep } \ov (2 \pi \ep )^n } 
\bgl\{ {1 \ov D} (\h g_\mn \h R_\rs + \h g_\rs \h R_\mn ) \nn \\ 
& - & {1 \ov 4} (\h g_{\m \r } \h R_{\n \s } + \h g_{\m \s } \h R_{\n \r } 
+ \h g_{\n \r } \h R_{\m \s } + \h g_{\n \s } \h R_{\m \r } ) 
\bgr\} \dt^D \ .  
\lbl{eqn:sin7} 
\eea
These correspond to the overlapping divergences, when a part of 
the $x$ derivatives of $G_0$'s is converted into those
with respect to $y$. 

%\vspace{0.5cm}

We also have to consider the products involving a function $G_1$. 
The equations (\ref{eqn:rel1}) make it possible for us to reduce the 
derivative of $G_1$ to the sum of $G_0$'s and $G_1$'s with no derivative. 
Using them, we find the singular parts of the products including 
a function $G_1$ and four derivatives to be proportional to the delta 
functions without any derivatives or curvature tensors. There are three 
types of products necessary for our calculation. 
\bea
& & \nab_\m \nab_\n G_0 \nab_\r \nab_\s G_1 \ct G_0{}^{n-1} \nn \\
& \sim & 
{(-1)^n \ov 2 D} {\m^{n \ep } \ov (2 \pi \ep )^n } 
\h g_\mn \h g_\rs \dt^D \nn \\  
& + & {1 \ov 2 (D+2)} {(-1)^n \ov n} {\m^{n \ep } \ov (2 \pi \ep )^n } 
\bgl( - {2 \ov D} \h g_\mn \h g_\rs + \h g_{\m \r } \h g_{\n \s } 
+ \h g_{\m \s } \h g_{\n \r } \bgr) \dt^D \ ,
\lbl{eqn:sin8} \\ 
& & \qq \nn \\ 
& & \nab_\m G_0 \nab_\n G_0 \nab_\r \nab_\s G_1 \nn \\
& \sim & 
- {1 \ov 4 D} {\m^{2 \ep } \ov (2 \pi \ep )^2 } 
\h g_\mn \h g_\rs \dt^D \nn \\  
& - & {\ep \ov 4 D (D+2)} {\m^{2 \ep } \ov (2 \pi \ep )^2 } 
(\h g_\mn \h g_\rs + \h g_{\m \r } \h g_{\n \s } 
+ \h g_{\m \s } \h g_{\n \r } ) \dt^D \ ,
\lbl{eqn:sin9} \\
& & \qq \nn \\ 
& & \nab_\m \nab_\n G_0 \nab_\r G_0 \nab_\s G_1 \nn \\
& \sim & 
- {1 \ov 4 (D+2)} {\m^{2 \ep } \ov (2 \pi \ep )^2 } 
\bgl( {2 \ov D} \h g_\mn \h g_\rs - \h g_{\m \r } \h g_{\n \s } 
- \h g_{\m \s } \h g_{\n \r } \bgr) \dt^D \ .
\lbl{eqn:sin10} 
\eea

%\vspace{0.5cm}

So far, we have considered the singular behavior of the products 
of $G_0$'s or $G_1$'s with only $x$ derivatives. Now, let us extend 
the previous results to the products with $x'$ derivatives as well as 
those with respect to $x$. 

We can convert the $x$ derivative of the functions $G_0 , G_1$ 
to that with respect to $x'$, using an identity which 
uses the parallel displacement bi-vector $\g_\m{}^{\n '}$:
\be
\g_{\m '}{}^\m \nab_\m G_i \ = \ - G_i {\ola \nab}_{\m '} \ , 
\ \ \ i=0,1. 
\ee
As a result, the singular expansions (\ref{eqn:sin1})-(\ref{eqn:sin5}) and 
(\ref{eqn:sin8})-(\ref{eqn:sin10}) are valid under the change:
\be
\nab_\m G_i \ \longrightarrow 
- \g_\m{}^{\m '} G_i {\ola \nab}_{\m '} \ , 
\ \ \ i=0,1. 
\ee
On the other hand, we should pay attention to the singular parts 
proportional to the quadratic derivatives of a delta function. 
In (\ref{eqn:sin6}) and (\ref{eqn:sin7}), we need additional curvature 
terms as 
\bea 
& & \g_\r{}^{\r '} \nab_\m G_0 {\ola \nab}_{\r '} \ct 
\g_\s{}^{\s '} \nab_\n G_0 {\ola \nab}_{\s '} \ct 
G_0{}^{n-1} \nn \\
& \sim & 
\nab_\m \nab_\r G_0 \nab_\n \nab_\s G_0 \ct G_0{}^{n-1} \nn \\
& - & {1 \ov D+2} {(-1)^n \ov n} {\m^{n \ep } \ov (2 \pi \ep )^n } 
\bgl\{ {1 \ov D} (\h g_{\m \r} \h R_{\n \s} + \h g_{\n \s} \h R_{\m \r} ) 
+ \h R_{\mn \rs} + \h R_{\m \s \r \n} \bgr\} \dt^D \ , 
\lbl{eqn:sin11} \\
& & \qq \nn \\ 
& & \g_\r{}^{\r '} \nab_\m G_0 {\ola \nab}_{\r '} \ct 
\nab_\n G_0 \ct \g_\s{}^{\s '} G_0 {\ola \nab}_{\s '} \ct 
G_0{}^{n-2} \nn \\
& \sim & 
\nab_\m \nab_\r G_0 \nab_\n G_0 \nab_\s G_0 \ct G_0{}^{n-2} \nn \\
& + & {\ep \ov 2 D (D+2)} {(-1)^n \ov n (n-1)} 
{\m^{n \ep } \ov (2 \pi \ep )^n } 
(\h R_{\mn \rs} + \h R_{\m \s \r \n} - \h g_{\n \s} \h R_{\m \r} ) 
\dt^D \ . 
\lbl{eqn:sin12}
\eea
Here, we have used the coincidence limits (\ref{eqn:lima}).

%\newpage
%\vspace{1cm}

\newpage

\addcontentsline{toc}{section}{\protect\numberline{}{Bibliography}}

\end{document}